\documentclass[%
 twocolumn,
nofootinbib,
 amsmath,amssymb,
 aps,
floatfix,
]{revtex4-1}
\usepackage[utf8x]{inputenc}
\usepackage{lmodern}
\usepackage[T1]{fontenc}
\usepackage{graphicx}
\usepackage{dcolumn}
\usepackage{bm}
\usepackage{verbatim}
\usepackage{url}
\usepackage{hyperref}
\usepackage{natbib}
\usepackage{etoolbox}
\usepackage{tikz}
\usepackage{longtable}

\newcommand{\tikzcircle}[2][red,fill=red]{\tikz[baseline=-0.5ex]\draw[#1,radius=#2] (0,0) circle ;}%
\newrobustcmd*{\tikzsquare}[1]{\tikz{\filldraw[draw=#1,fill=#1] (0,0) rectangle (0.6em,0.6em);}}%
\newrobustcmd*{\tikztriangle}[1]{\tikz{\filldraw[draw=#1,fill=#1] (0,0) -- (0.2cm,0) -- (0.1cm,-0.2cm);}}

\begin{document}

\title{Deuteron and Antideuteron Production Simulation in Cosmic-ray Interactions}

\author{Diego-Mauricio Gomez-Coral}
\email{diegomez@estudiantes.fisica.unam.mx}
\author{Arturo Menchaca Rocha}
\author{Varlen Grabski}
\affiliation{%
Instituto de Física, Universidad Nacional Autónoma de México\\
Circuito de la Investigación Científica, Ciudad de México, México\\
}%

\author{Amaresh Datta}
\author{Philip von Doetinchem}
\author{Anirvan Shukla}
\affiliation{%
Department of Physics and Astronomy, University of Hawaii at Manoa\\
2505 Correa Rd, Honolulu, HI 96822, USA\\
}%

\date{\today}

\begin{abstract}
The study of the cosmic-ray deuteron and antideuteron flux receives an increasing interest in current astrophysics investigations. For both cases an important contribution is expected from the nuclear interactions of primary cosmic rays with intergalactic matter. In this work, deuteron and antideuteron production from 20 to 2.6$\times$10$^{7}$\,GeV beam energy in p+p and p+A collisions were simulated using EPOS-LHC and Geant4's FTFP-BERT Monte Carlo models by adding an event-by-event coalescence model afterburner. These estimates depend on a single parameter ($p_0$) obtained from a fit to the data. The $p_0$ for deuterons in this wide energy range was evaluated for the first time. It was found that $p_0$ for antideuterons is not a constant at all energies as previous works suggested and as a consequence the antideuteron production cross section can be at least 20 times smaller in the low collision energy region, than earlier estimations.
\end{abstract}

\maketitle

\section{Introduction}

Deuteron abundance measurements in cosmic rays (CRs) \cite{caprice, tomassetti} have shown that cosmic deuteron formation is understood as the result of the nuclear interactions of primary CRs, mainly protons and helium, with the interstellar media (ISM) also composed mostly of H and He. This cosmic deuteron source, known as secondary production, is dominated by two contributions: fragmentation of CRs nuclei ($^3$He, and $^4$He) with the hydrogen and helium from the ISM, and the resonant reaction $p+p\rightarrow d+\pi^{+}$, in which deuterons are produced in a narrow energy distribution (FWHM $\approx$ 320\,MeV) with the maximum around $\sim$ 600\,MeV \cite{meyer}. This last reaction is only significant for energies below 1\,GeV meanwhile fragmentation is the main origin for deuterons at higher energies. As a consequence, the cosmic deuteron flux provides important information about CRs propagation in the Galaxy, such as the mean amount of ISM that primary CRs encounter as they travel from their sources to the Earth. 

Besides the two processes described above, accelerator experiments revealed a third deuteron production mechanism, explained within the framework of the so-called coalescence model \cite{kapusta, butler, baltz, Schwarzschild}. This applies to free nucleons resulting from CRs-ISM interactions, in which residual protons and neutrons lie sufficiently close in phase space to form deuterons. Such free nucleons may be the result of p+nuclei fragmentation interactions. At sufficiently high energies, p+p and p+nuclei interactions can also create multiple nucleon-antinucleon pairs, generating conditions for the formation of deuterons through the coalescence mechanism, not incorporated yet in the standard calculation of the secondary deuteron CRs flux.

Note that, of the three deuteron-producing mechanisms described above, coalescence is the only one that also allows the formation of secondary antideuterons. The secondary antideuteron flux is predicted to have a maximum at a kinetic energy per nucleon T $\approx$ 4\,GeV/n, and to fall sharply at lower T values \cite{Chardonnet1, duperray_flux_2005, Ibarra:2013qt}. This is interesting because a number of dark matter models suggest an antideuteron flux from dark matter annihilation or decays to be about two orders of magnitude higher than the secondary background at energies of about 1\,GeV/n \cite{vonDoetinchem}. Hence, the predicted low energy secondary antideuteron-suppressed window has generated great interest in dark matter research \cite{Donato1, Donato2, Ibarra2, Fornengo, baer, profumo, barrau}, stimulating the experimental exploration for cosmic antideuterons. Currently the Alpha Magnetic Spectrometer experiment (AMS-02) on board of the International Space Station is searching for cosmic antideuterons, and in the near future the balloon borne General Antiparticle Spectrometer (GAPS) will join in that quest. As detectors sensitivity increases and observational limits are set, a precise calculation of the secondary antideuteron flux is more important, including additional antideuteron background sources like those represented by the detection instruments and the atmosphere above them.

The aim of this study is to benefit from the continuous improvement of Monte Carlo (MC) particle interaction simulators as well as the development of an afterburner\footnote[1]{Name given to routines commonly used in MC codes to modify the particle distribution produced by the generator according to a model.} for (anti)deuteron coalescence. This tool allows to perform predictions about the deuteron and antideuteron production, consistent with available accelerator data from a wide energy range. Section~\ref{s-cm} reviews the coalescence model, as well as the approximations used by previous authors to predict (anti)deuteron production. In section~\ref{s2} the available proton and antiproton data from accelerator experiments are compared to MC models with the aim to define which generator provides the best results over the energy range of interest. In section~\ref{s-d}, the implementation of the afterburner to produce $\text{d}$ and $\bar{\text{d}}$ in an event-by-event approach is described. Deuteron and antideuteron measurements are fitted with simulations using the afterburner to determine the best coalescence momentum parameter. Conclusions are presented in Section~\ref{s-c}.

\section{Coalescence Model\label{s-cm}}

To describe (anti)deuteron formation we use the coalescence model \cite{kapusta, butler, baltz}. This postulates that proton-neutron (pn) or antiproton-antineutron pairs ($\bar{\text{p}}\bar{\text{n}}$) that are close enough in phase space could result in the formation of deuterons ($\text{d}$) or antideuterons ($\bar{\text{d}}$), respectively. In the remaining of this section the antinucleon notation will be used, although the equations are equally valid for nucleons. This formation occurs with a probability $C(\sqrt{s}, \vec{k}_{\bar{p}}, \vec{k}_{\bar{n}})$, known as the coalescence function. $C$ depends on the momentum difference $2\Delta \vec{k}=\vec{k}_{\bar{p}}-\vec{k}_{\bar{n}}$ and on the total energy available ($\sqrt{s}$). Following the derivation presented in \cite{Donato1, Fornengo}, the momentum distribution of antideuterons produced in the coalescence scheme can be expressed as:

\begin{multline}\label{s2:eq1}
\scriptsize
\left( \frac{dN_{\bar{d}}}{d\vec{k}_{\bar{d}}^3} \right) (\sqrt{s}, \vec{k}_{\bar{d}}) = \int {d^3\vec{k}_{\bar{p}}} {d^3\vec{k}_{\bar{n}}} \times \\
\left( \frac{dN_{\bar{p}\bar{n}}}{d\vec{k}_{\bar{p}}^{3} d\vec{k}_{\bar{n}}^{3}}(\sqrt{s}, \vec{k}_{\bar{p}}, \vec{k}_{\bar{n}}) \right) C(\sqrt{s}, \vec{k}_{\bar{p}}, \vec{k}_{\bar{n}}) \delta (\vec{k}_{\bar{d}}-\vec{k}_{\bar{p}}-\vec{k}_{\bar{n}})
\end{multline}

Where $dN_{\bar{d}}=d^{3} \sigma_{\bar{d}}/\sigma_{tot}$, with $\sigma_{tot}$ and $d^{3} \sigma_{\bar{d}}$ being the total and differential cross sections and $dN_{\bar{p}\bar{n}}=d^{6} \sigma_{\bar{p}\bar{n}}/\sigma_{tot}$ the number of pairs ($\bar{\text{p}}\bar{\text{n}}$) produced in the collision. 
 
As a first approximation, it is assumed that the coalescence function does not depend on collision energy, resulting in $C(\sqrt{s}, \Delta \vec{k}) = C(\Delta \vec{k})$. Next, $C$ is approximated by a step function $\Theta(\Delta k^2-p_{0}^2)$ where $p_0$ is a free parameter called the coalescence momentum, representing the magnitude of the maximal radius in momentum space that allows antideuteron formation. Under this approximation, the probability changes from zero when $|\Delta \vec{k}| > p_{0}$ to one if $|\Delta \vec{k}| < p_{0}$. After a convenient variable transformation, and considering that $|\Delta \vec{k}| \ll |\vec{k}_{\bar{d}}|$, Eq.\,(\ref{s2:eq1}) becomes: 

\begin{multline}\label{s2:eq2}
\scriptsize
\gamma_{\bar{d}} \left( \frac{d N_{\bar{d}}}{d\vec{k}_{\bar{d}}^3} \right) (\sqrt{s}, \vec{k}_{\bar{d}})  \simeq \left[ \frac{4\pi p_0^3}{3} \right] \\
\times \gamma_{\bar{p}} \gamma_{\bar{n}} \left( \frac{d N_{\bar{p}\bar{n}}}{d\vec{k}_{\bar{p}}^{3} d\vec{k}_{\bar{n}}^{3}}(\sqrt{s}, \vec{k}_{\bar{p}}=\vec{k}_{\bar{d}}/2, \vec{k}_{\bar{n}}=\vec{k}_{\bar{d}}/2) \right)
\end{multline}

\begin{figure*}[!hbt]
\begin{center}
\begin{tabular}{l l}
\includegraphics[width=8.8cm]{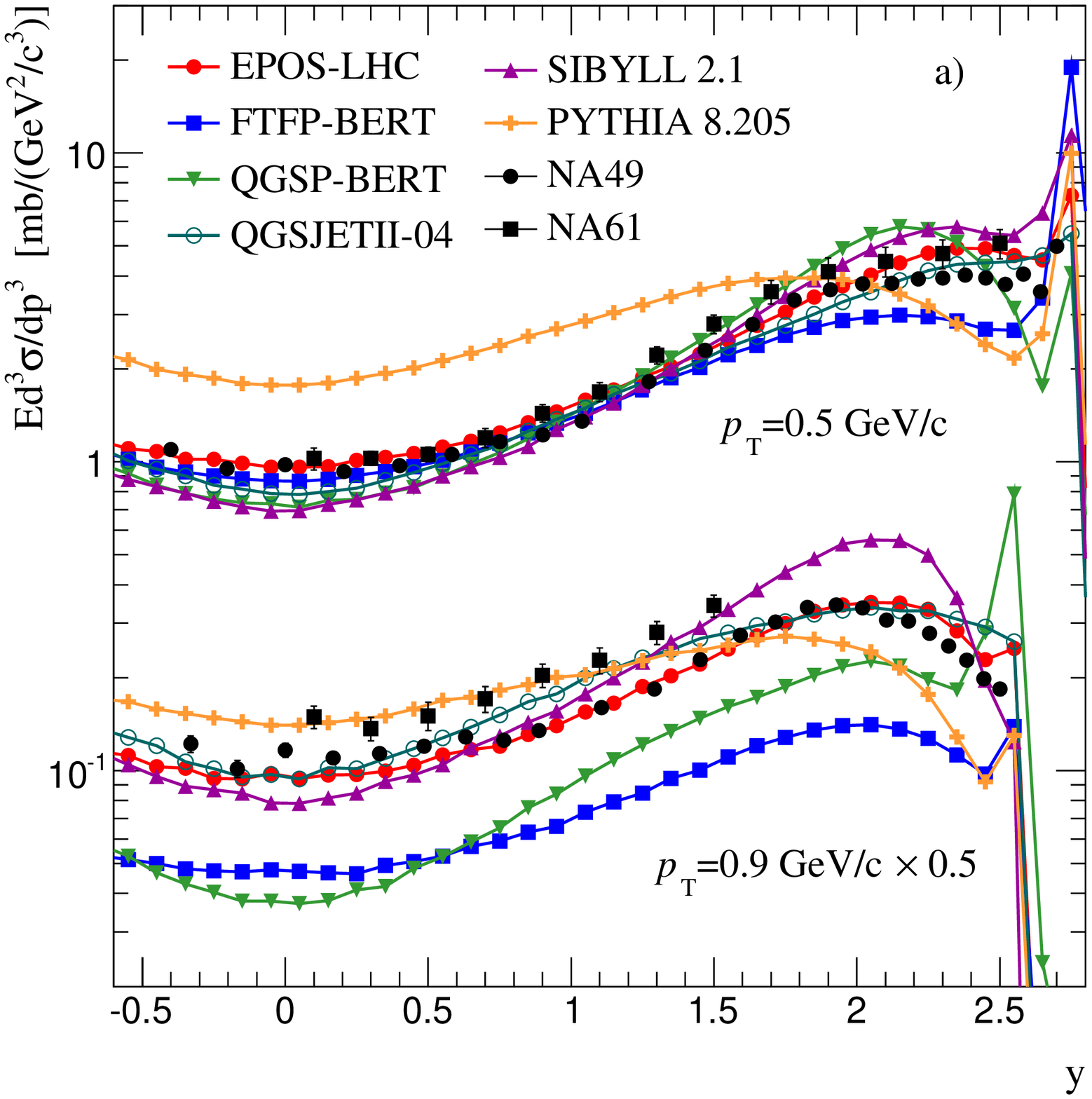} &
\includegraphics[width=8.8cm]{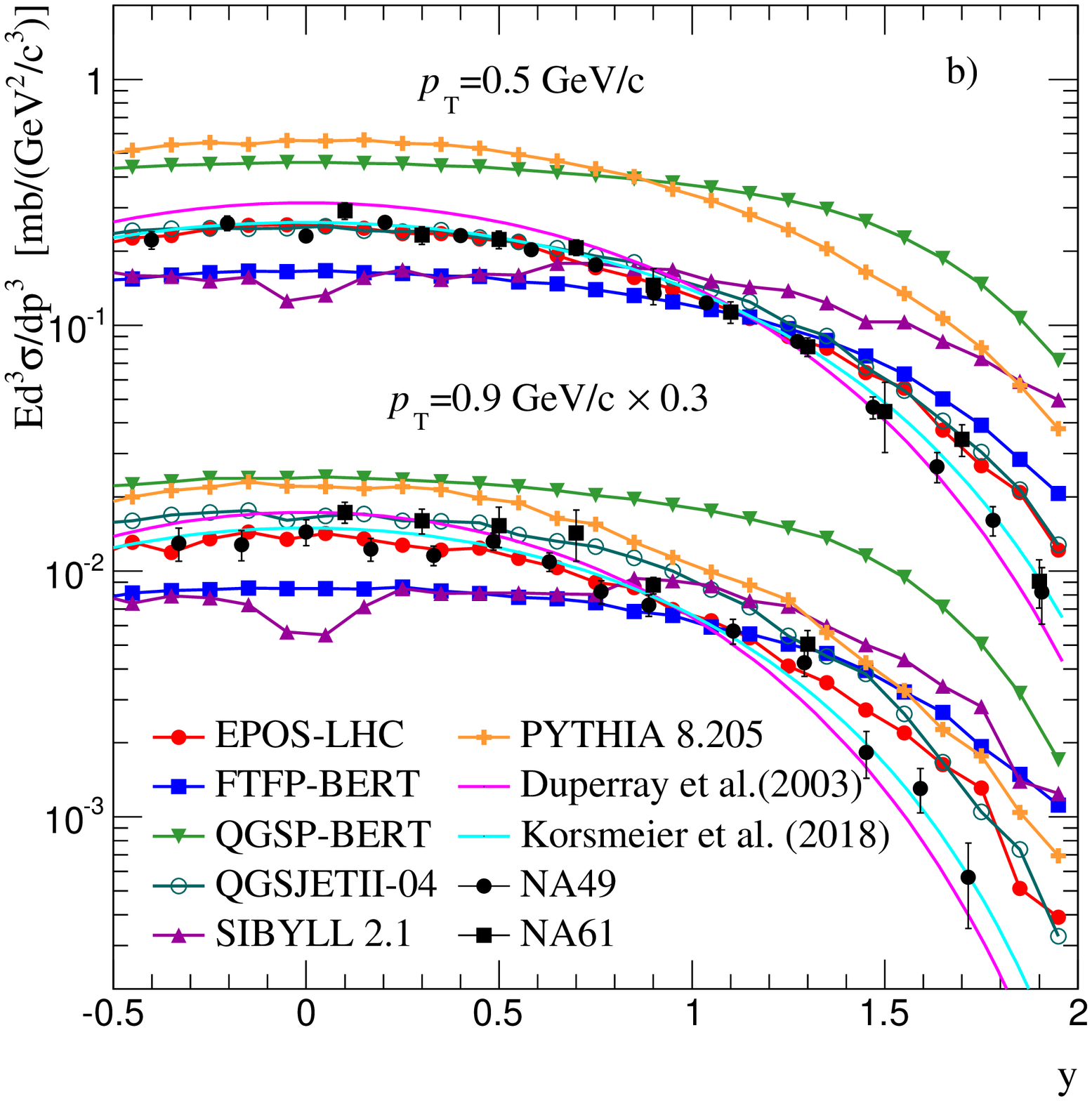}
\end{tabular}
\caption{(Color online) Invariant differential cross sections as function of rapidity ($y$) are calculated with different MC models for protons \textbf{a)}, and antiprotons \textbf{b)} in p+p collisions at 158\,GeV/$c$. Results for two bins of transverse momentum ($p_T$) are compared with data from experiments NA49 \cite{Anticic:2009wd} and NA61 \cite{na61_2017}.}
\label{s2:fig0}
\end{center}
\end{figure*} 

Where the $\gamma$ factor was introduced to show the result in a Lorentz-invariant form. Eq.\,(\ref{s2:eq2}) indicates that antiproton and antineutron momentum distributions as well as the coalescence momentum are necessary to estimate the antideuteron cross section. Assumptions of independent (uncorrelated) production of antiprotons and antineutrons have been used in analytical calculations \cite{Chardonnet1}, to express the momentum distribution of the pair ($ dN_{\bar{p}\bar{n}}/d\vec{k}_{\bar{p}}^{3} d\vec{k}_{\bar{n}}^{3}$) as the product of two independent isotropic distributions ($ dN_{\bar{p}}/d\vec{k}_{\bar{p}}^{3} \times dN_{\bar{n}}/d\vec{k}_{\bar{n}}^{3}$). This is known as the analytical coalescence model. This assumption, however, is overly simplistic \cite{Kadastik, Ibarra:2013qt, Ibarra2} since correlations have an important effect on deuteron and antideuteron formation. MC generators take into account the correlations involved in the production with the caveat that there can be uncertainties in the description of correlation effects. Such effects may be related to phase space availability, spin alignments, energy conservation, antiproton-antineutron production asymmetry etc. These possible effects are absorbed in the coalescence momentum $p_{0}$.

\begin{table*}[!htb]
\begin{ruledtabular}
\begin{tabular}{  c  c  c  c  c  c  c }                          
   \textbf{Experiment or} & \textbf{Reference} & \textbf{Collision} &  \textbf{Final states} 	& \boldmath{$p_{lab}$} & \boldmath{$\sqrt{s}$}  &  \textbf{Phase Space}\\
	\textbf{Laboratory}		&				&				&			&	\textbf{(GeV/\boldmath{$c$})}		&	\textbf{(GeV)}		&	 \\						
 \hline
   	ITEP \footnotetext[1]{No feed-down correction} \footnotemark[1]		&\cite{Vorontsov}				&  p+Be	& p		&   10.1& 4.5	&	1$\leq p\leq$7.5\,GeV/$c$; $\theta$ = 3.5 deg\\

   CERN	\footnotemark[1]	&\cite{Allaby1970, diddens_particle_2008} &  p+p& p, $\bar{\text{p}}$ & 19.2	& 6.1	&	2$\leq p\leq$19\,GeV/$c$; \\
				&								&  p+Be 		& p, $\bar{\text{p}}$ &		&		&	0.72 $\leq \theta \leq$ 6.6 deg\\

	CERN \footnotemark[1]		&\cite{diddens_particle_2008}	&  p+p			& p	 &  24	& 6.8	&	2$\leq p\leq$9\,GeV/$c$; $\theta$ = 6.6 deg\\

	NA61/SHINE	&\cite{Abgrall2016}				&  p+C			& p  &  31	& 7.7	&	0$\leq p\leq$25\,GeV/$c$; 0$\leq \theta \leq$ 20.6 deg\\
				&\cite{na61_2017}				&  p+p			& p, $\bar{\text{p}}$ &   &   &	$p_{T} \leq$ 1.5\,GeV/$c$; 0.1$\leq y\leq$2.0\\

	NA61/SHINE	&\cite{na61_2017}				&  p+p			& p, $\bar{\text{p}}$ & 40  &  8.8	&	$p_{T} \leq$ 1.5\,GeV/$c$; 0.1$\leq y\leq$2.0\\

   	Serpukhov \footnotemark[1]	&\cite{abramov_production_1980, Abramov_Baldin1985}
												&  p+p			& p, $\bar{\text{p}}$ &	70	& 11.5	&	0.48$\leq p_T\leq$ 4.22\,GeV/$c$; $\theta_{lab}$ = 9.2 deg\\
				&\cite{Abramov_Baldin1987}		&  p+Be 		& p, $\bar{\text{p}}$ &		&		&	\\
				&\cite{Abramov:1983es}			&  p+Al 		& p, $\bar{\text{p}}$ &		&		&	\\
   
	NA61/SHINE	&\cite{na61_2017}				&  p+p			& p, $\bar{\text{p}}$ & 80  & 12.3  &	$p_{T} \leq$ 1.5\,GeV/$c$; 0.1$\leq y\leq$2.0\\

   CERN-NA49	&\cite{Anticic:2009wd}			&  p+p			& p, $\bar{\text{p}}$ & 158	& 17.5	&	$p_{T} \leq$ 1.9\,GeV/$c$; $x_{F} \leq$1.0\\
				&\cite{Baatar:2012fua}			&  p+C			& p, $\bar{\text{p}}$ &		&	\\

   CERN-NA61	&\cite{na61_2017}				&  p+p			& p, $\bar{\text{p}}$ & 	& 	&	$p_{T} \leq$ 1.5\,GeV/$c$; 0.1$\leq y\leq$2.0\\
   
   CERN-SPS \footnotemark[1]		&\cite{BOZZOLI1978317, Baker:1974fv} &  p+Be & p, $\bar{\text{p}}$ &  200 & 19.4 & 23$\leq p\leq$197\,GeV/$c$	\\
   				&									 &  p+Al & p, $\bar{\text{p}}$ &	  &		 &	$\theta_{lab}$ = 3.6 mr, $\theta_{lab}$ = 0	\\

   Fermilab \footnotemark[1]  	&\cite{Antreasyan, Cronin}		&  p+p	& p, $\bar{\text{p}}$ &	300 & 23.8	&	0.77 $\leq p_T\leq$ 6.91\,GeV/$c$; \\
   			  	&								&  p+Be	& p, $\bar{\text{p}}$ &		& 		&	$\theta_{lab}$ = 4.4 deg, $\theta_{cm}$ = 90 deg\\

   Fermilab \footnotemark[1]  	&\cite{Antreasyan, Cronin}		&  p+p	& p, $\bar{\text{p}}$ &	400 & 27.4	&	0.77 $\leq p_T\leq$ 6.91\,GeV/$c$; $\theta_{lab}$ = 4.4 deg\\
   			  	&								&  p+Be	& p, $\bar{\text{p}}$ &  	    &    	&	\\

   CERN-ISR 	&\cite{Alper1975}				&  p+p	& p, $\bar{\text{p}}$ & 1078 & 45.0 &	0.1$< p_{T} < $4.8\,GeV/$c$; 0.0$\leq y \leq$1.0\\

   CERN-ISR 	&\cite{Alper1975}				&  p+p	& p, $\bar{\text{p}}$ & 1498 & 53.0 &	0.1$< p_{T} < $4.8\,GeV/$c$; 0.0$\leq y \leq$1.0\\
 
   CERN-LHCb	&\cite{lhcb}	&  p+He	& $\bar{\text{p}}$ & 6.5$\times$ 10$^{3}$ & 110	&	0.0$\leq p_{T}\leq$4.0\,GeV/$c$; 12$\leq p\leq$110\\

   CERN-ALICE	&\cite{aamodt_production_2011}	&  p+p	& p, $\bar{\text{p}}$ & 4.3$\times$ 10$^{5}$ & 900	&	0.0$\leq p_{T}\leq$2.0\,GeV/$c$; -0.5$\leq y\leq$0.5\\
 
   CERN-ALICE	&\cite{aamodt_production_2011}	&  p+p  & p, $\bar{\text{p}}$ & 2.6$\times$ 10$^{7}$ & 7000	&	0.0$\leq p_{T}\leq$2.0\,GeV/$c$; -0.5$\leq y\leq$0.5\\

\end{tabular}
\end{ruledtabular}
\caption{List of experimental data on proton and antiproton production in p+p and p+A collisions considered in this work to compare with simulations.}
\label{s2:tab1}
\end{table*}

\section{$\text{p}$ and $\bar{\text{p}}$ Production Simulation}\label{s2}

To produce (anti)deuterons using MC generators, it is necessary to have a correct prediction of the (anti)proton production. In the present study high energy MC generators have been preferred over their counterparts at low energy. Our choice is based on the conclusions presented in reference \cite{Kachelriess}, where the authors showed that MC models used in low energy nuclear physics have strong deviations (up to an order of magnitude) from the measured $\bar{\text{p}}$ spectra, while they demonstrate that advanced high energy MC generators like EPOS-LHC \cite{Pierog:2013ria} predict reliably the antiproton yield. Furthermore, these generators have been tuned to experimental results in a wide energy range, and they are extensively and consistently used in simulating CRs interactions.

Here, several MC models were tested and compared to (anti)proton data. An example is shown in Fig.\,\ref{s2:fig0}, where the Cosmic Ray Monte Carlo package (CRMC) \cite{crmc} was used to estimate invariant differential cross sections as a function of rapidity ($y$) using EPOS-LHC \cite{Pierog:2013ria}, QGSJETII-04 \cite{Ostapchenko:2004qz}, and SIBYLL2.1 \cite{Ahn:2009wx}. The figure also includes the predictions of PYTHIA-8.205 \cite{Sjostrand:2007gs} and two Geant4 (version:10.02.p02) \cite{Agostinelli2003250} hadronic models: FTFP-BERT (based on the Fritiof description of string fragmentation \cite{fritiof} with the Bertini intra-nuclear cascade model) and QGSP-BERT (quark-gluon string based model \cite{qgsp} with the Bertini intra-nuclear cascade model).

\begin{figure}[!hbt]
\begin{center}
\includegraphics[width=8.6cm]{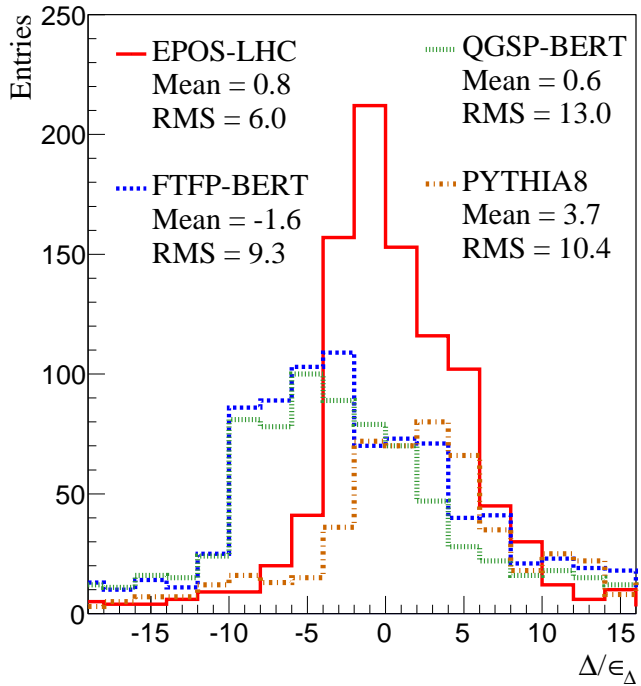}
\caption{(Color online) Distributions of the difference between measurements and the MC generators divided by the error (see Eq.\,\ref{s3:eq1}) for proton production in p+p and p+A collisions.}
\label{s2:fig1}
\end{center}
\end{figure} 

In Table\,\ref{s2:tab1} a list of the experimental data considered in this work is shown along with their collision characteristics. The selection of these experimental data was based on their relevance to the most abundant cosmic ray species, as well as to the energy range in which deuterons and antideuterons are produced in CRs collisions. Since part of the available experimental data is old enough to lack the precision tracking and vertex determination techniques available today, this might have introduced inherent systematic uncertainties. For example, feed-down contribution to protons and antiprotons (from decays of heavier baryons) were not handled well in some of these data, contributing to the mismatch between data and MC production. The detected fraction of protons and antiprotons produced by this mechanism depends on the energy boost generated by the parent hyperons decay, as well as the details of the detector. This makes it difficult to estimate, \textit{a posteriori}, the proper correction \cite{KapplandWinkler, winkler, annikawinkler}. For the case of experiments at CERN-ISR, where p+p collisions with center of mass energy from 23 to 53\,GeV were studied, a correction was possible. According to \cite{Anticic:2009wd}, the detector design of this experiment allowed nearly all baryonic decay products to be included in the measured cross section. Thus, here the corresponding correction factors were extracted from simulations and applied to this group of data. This was not the case for other data sets, as indicated in Table\,\ref{s2:tab1}. 

To determine which MC is describing (anti)proton measurements most reliably in the energy range considered, a quantitative comparison between MC models, parametrizations and data is made with the help of Eq.\,\ref{s3:eq1}.

\begin{equation}\label{s3:eq1}
\frac{\Delta}{\epsilon_{\Delta}} = \frac{ \left ( E\frac{d^3 \sigma}{dp^3}^{sim}-E\frac{d^3 \sigma}{dp^3}^{data} \right ) }{\sqrt{(\epsilon_{sim})^2+(\epsilon_{data})^2}}
\end{equation}

This equation allows to calculate the difference ($\Delta$) between measurement and simulated differential cross sections ($Ed^3\sigma/dp^3$). Then $\Delta$ is divided by the total error ($\epsilon_{\Delta}$). The resulting quantity ($\Delta/\epsilon_{\Delta}$) is evaluated for every data set listed in Table \ref{s2:tab1}, and their distributions for a choice of models are illustrated in Figs.\,\ref{s2:fig1} and \ref{s2:fig3} for protons and antiprotons, respectively. The rest of the models are compared in appendix \ref{appendix1} (Figs.\,\ref{ap0:fig1} and \ref{ap0:fig2}). Ideally, these distributions should be centered at zero with the RMS value close to 1 when the measurement and the theoretical value are compatible on an absolute scale.

\begin{figure}[h]
\begin{center}
\includegraphics[width=8.6cm]{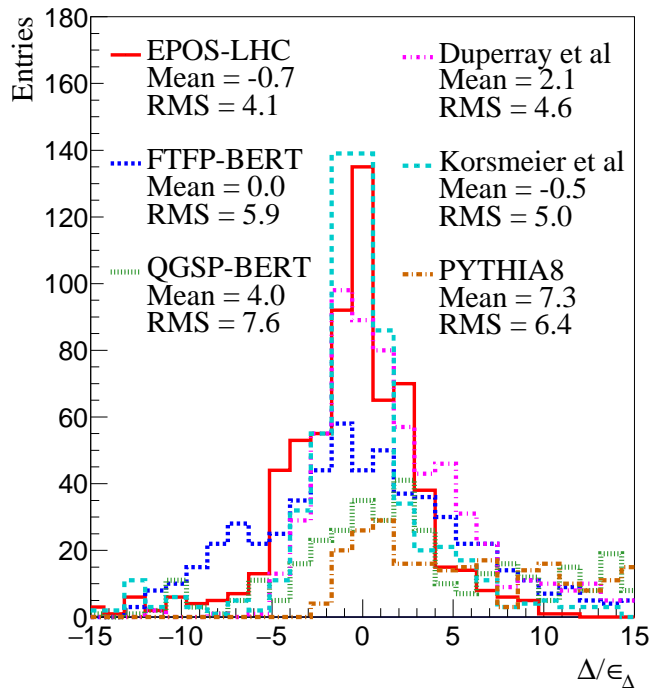}
\caption{(Color online) Distributions of the difference between measurements and the MC generators divided by the error (see Eq.\,\ref{s3:eq1}) for antiproton production in p+p and p+A collisions.}
\label{s2:fig3}
\end{center}
\end{figure}

\begin{table*}[!hbt]
\begin{ruledtabular}
\begin{tabular}{  c  c  c  c  c  c  c  c}                           
    \textbf{Experiment or} & \textbf{Reference} & \textbf{Collision} & \textbf{$p_{lab}$} & \textbf{$\sqrt{s}$}  & \multicolumn{2}{c}{No. of points} & \textbf{Phase Space}\\
	\cline{6-7}
	\textbf{Laboratory}		&					&					&	(GeV/$c$)		&	(GeV)	& $\text{d}$	& $\text{dbar}$ &	\\
 \hline
	CERN		&\cite{diddens_particle_2008}			&  p+p	&  19	& 6.15	& 6	& 0	& 0$\leq p\leq$9\,GeV; $\theta$ = 6.6 deg\\
	CERN		&\cite{diddens_particle_2008}			&  p+p	&  24	& 6.8	& 4	& 0	& 0$\leq p\leq$9\,GeV; $\theta$ = 6.6 deg\\					
   Serpukhov 	&\cite{Abramov_Baldin1987} 				&  p+p	&  70	& 11.5	& 7	& 2	& 0.48$\leq p_T\leq$ 2.4\,GeV; $\theta_{lab}$ = 9.2 deg\\
				&										&  p+Be &		&		& 6	& 3	& \\   
   CERN-SPS		&\cite{Bussiere:1980yq,BOZZOLI1978317}	&  p+Be	&  200	& 19.4	& 3	& 5	& 15$\leq p_{lab}\leq$ 40\,GeV; $\theta_{lab}$ = 0 deg\\
   				&										&  p+Al	&  		&		& 3	& 3	& \\
   Fermilab  	&\cite{Cronin}							&  p+Be	&  300 	& 23.8	& 4	& 1	& 0.77 $\leq p_T\leq$ 6.91\,GeV; $\theta_{lab}$ = 4.4 deg\\

   CERN-ISR 	&\cite{Alper2, gibson_production_2008, albrow}	&  p+p	&  1497.8 & 53 	& 3	& 8	& 0.0$\leq p_{T}\leq$1.0 ; $\theta_{cm}$ = 90 deg\\
 
   CERN-ALICE	&\cite{eulogio, eulogio_paper}						&  p+p	&  4.3$\times$ 10$^{5}$	& 900	& 3	& 3	& 0.0$\leq p_{T}\leq$2.0 ; -0.5$\leq y\leq$0.5\\
  
   CERN-ALICE	&\cite{eulogio, eulogio_paper, alicedeuteron}							&  p+p	&  2.6$\times$ 10$^{7}$ & 7000	& 21 & 20 & 0.0$\leq p_{T}\leq$2.0 ; -0.5$\leq y\leq$0.5\\
  
\end{tabular}
\end{ruledtabular}
\caption{List of experimental data on deuteron and antideuteron production in p+p and p+A collisions considered in this work to compare with simulations.}
\label{s3:tab1}
\end{table*}

Fig.\,\ref{s2:fig1} illustrates how proton production in p+p and p+A collisions is in general better described by EPOS-LHC. Yet, the corresponding distribution shows a positive-value tail. The origin of these deviations as function of the collision momenta are described also in appendix \ref{appendix1}. A similar analysis for antiprotons is presented in Fig.\,\ref{s2:fig3}, but in this case we added the parameterization of Duperray \textit{et al.} \cite{Duperray_2003} and the parametrization presented by Winkler \cite{winkler} which was updated by Korsmeier \textit{et al.} \cite{Korsmeier} to the latest NA61 and LHCb data. As in the case of protons, the antiproton prediction from EPOS-LHC provides better results than other MC models, while being comparable to the parametrizations. The dependence of the positive and negative value tail of EPOS-LHC in Fig.\,\ref{s2:fig3} with the collision momenta are described in appendix \ref{appendix1}.

From the results shown above, the EPOS-LHC estimates for proton and antiproton production would be the natural choice. Yet, because the Geant4 framework is broadly used in simulations of particle interactions with detectors, here the Geant4 hadronic model FTFP-BERT predictions are also included. Note however, the use of this MC model is limited to a kinetic energy collision T\,$<$\,10\,TeV.

\section{$\text{d}$ and $\bar{\text{d}}$ Production Simulation\label{s-d}}

\subsection{Estimation of Coalescence Momentum}\label{s4}

To generate (anti)deuterons emulating the coalescence process, an afterburner \cite{eulogio} was created to be coupled to the MC generators EPOS-LHC and FTFP-BERT. The afterburner performed an iterative operation for every event, by identifying all proton-neutron and antiproton-antineutron pairs from the stack of particles created by the generator and calculating the difference in momenta of each pair in their center-of-mass frame. Half of the magnitude of this difference ($\Delta k=|\vec{k}_{\bar{p}}-\vec{k}_{\bar{n}}|/2$) was compared to the coalescence momentum $p_0$. If $\Delta k$ was lower than $p_0$, (an)a (anti)deuteron with momentum $\vec{k}_{d}=\vec{k}_{p}+\vec{k}_{n}$ (or $\vec{k}_{\bar{d}}=\vec{k}_{\bar{p}}+\vec{k}_{\bar{n}}$) and energy $E_d=\sqrt{\vec{k}_{d}^2+m_{d}^2}$ (or $E_{\bar{d}}=\sqrt{\vec{k}_{\bar{d}}^2+m_{\bar{d}}^2}$) was included in the stack, while the corresponding nucleons were deleted from it. (Anti)protons and (anti)neutrons from weak decays were excluded from the simulations. 

\begin{figure}[!ht]
\begin{center}
\includegraphics[width=8.6cm]{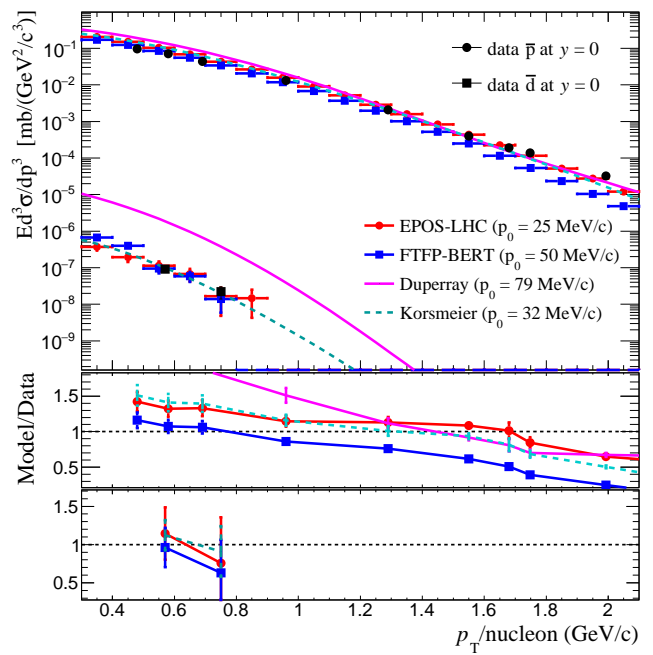}
\caption{(Color online) Antiproton and antideuteron invariant differential cross sections in p+p collisions at 70\,GeV/$c$ as function of transverse momentum ($p_T$) calculated with EPOS-LHC, FTFP-BERT and parametrizations \cite{Duperray_2003, Korsmeier}. The results are compared to data \cite{abramov_production_1980, Abramov_Baldin1985, Abramov_Baldin1987} (see text for details).}
\label{s4:fig4}
\end{center}
\end{figure}

\begin{figure*}[!htb]
\begin{center}
\begin{tabular}{ll}
\includegraphics[width=8.6cm]{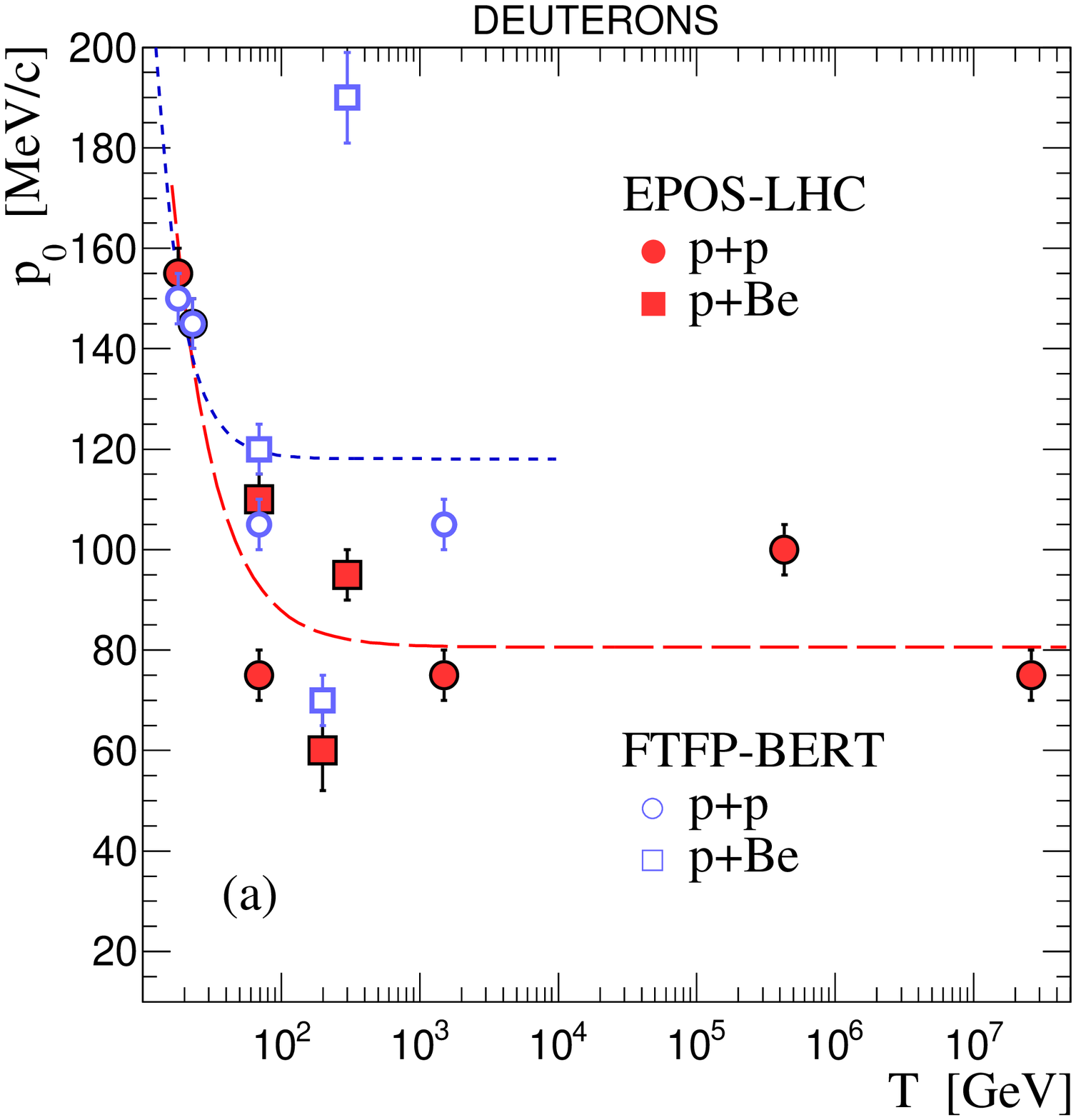}
& \includegraphics[width=8.6cm]{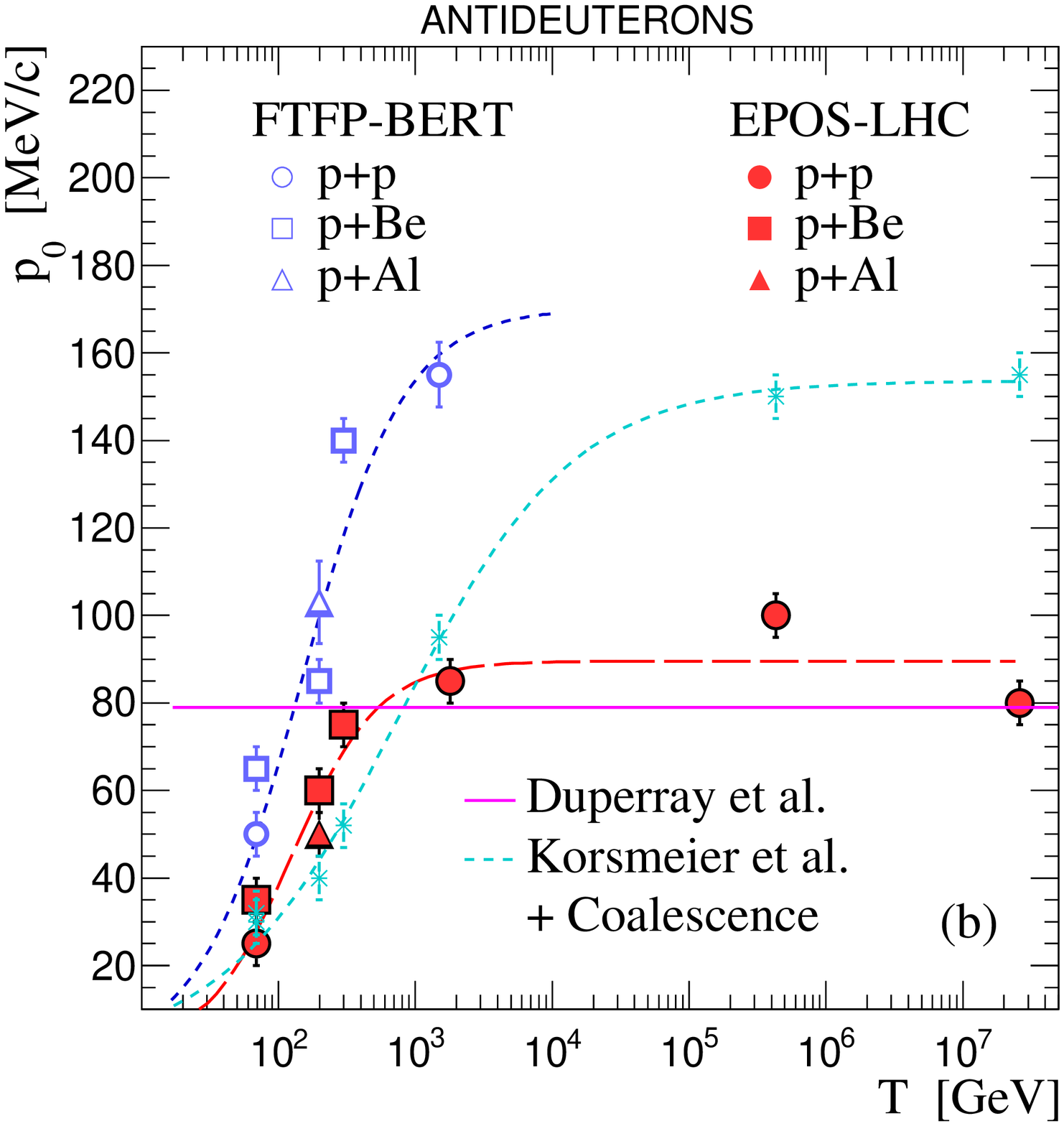}\\
\end{tabular}
\caption{(Color online) Extracted coalescence momentum $p_0$ (symbols) for deuterons \textbf{(a)} and  antideuterons \textbf{(b)} as function of the collision kinetic energy (T). Fit functions [Eqs.\,(\ref{s4:eq1}) and (\ref{s4:eq2})] for EPOS-LHC (long-dashed red line) and FTFP-BERT (dashed blue line) are shown. Additionally, the $p_0$ values obtained from the analytic coalescence model and the parametrization of Korsmeier \textit{et al.} are included (dashed cyan line and dots). Also, the constant value of $p_0=79$\,MeV/$c$ estimated by Duperray \textit{et al.} is plotted (solid magenta line).}
\label{s4:fig5}
\end{center}
\end{figure*}

The coalescence momentum was varied in steps of 5\,MeV/$c$, and the (anti)deuteron spectra corresponding to each of these values were compared with the experimental data in Table \ref{s3:tab1}. The $p_0$ that produced the lowest $\chi^2$ fit was thus selected. As an example of the results from this analysis, in Fig.\,\ref{s4:fig4} the p+p at 70\,GeV/$c$ case is presented. As observed, the best values of $p_0$ at this particular energy were 25\,MeV/$c$ for EPOS-LHC and 50\,MeV/$c$ for FTFP-BERT. In the Korsmeier \textit{et al.} parametrization case, $p_0$ was evaluated using the analytical expression in Eq.\,\ref{s2:eq2} assuming antiproton-antineutron independence and symmetry (i.e., the analytical coalescence model), which was fitted to data resulting in a $p_0=\text{32\,MeV/}c$ (cyan broken line in Fig.\,\ref{s4:fig4}). Duperray \textit{et al.} proposed a constant $p_0=\text{79\,MeV/}c$ over the whole energy range, also shown in Fig.\,\ref{s4:fig4} (magenta solid line).

The differential cross sections computed with the resulting $p_0$ values for EPOS-LHC, FTFP-BERT, as well as the parameterizations \cite{Duperray_2003, Korsmeier} are compared with the data in appendix\,\ref{appendix}. The values of $p_0$ extracted from the comparison to data are shown in Fig.\,\ref{s4:fig5} (a) for deuterons and in Fig.\,\ref{s4:fig5} (b) for antideuterons, as function of the collision kinetic energy (T) in the laboratory system. Although the trend of the $p_0$ values obtained with different MC models as a function of T is similar, their magnitude differ from one simulator to the other and also with respect to the parametrizations. Differences between MC models and parametrizations result from the correlations (or anticorrelations) in the antinucleon pairs only present in the MC generators \cite{Ibarra:2013qt, Fornengo, Ibarra2, Kadastik}. Disparities in the corresponding MC model assumptions, lead to deviations of their predictions for nucleon and antinucleon production, causing differences in the extracted $p_0$ among MC generators. To compare the coalescence momentum among MC models it is useful to factorize the (anti)nucleon mismatch assuming uncorrelated and symmetric production, hence treating the $p_0$ difference as due to antiproton mismatch. The details and results of this process are shown in appendix \ref{appendix2}. As shown in the next section this factorization however, has no effect on the deuteron and antideuteron cross section calculations.

\begin{figure*}[!htb]
\begin{center}
\begin{tabular}{ll}
\includegraphics[width=8.6cm]{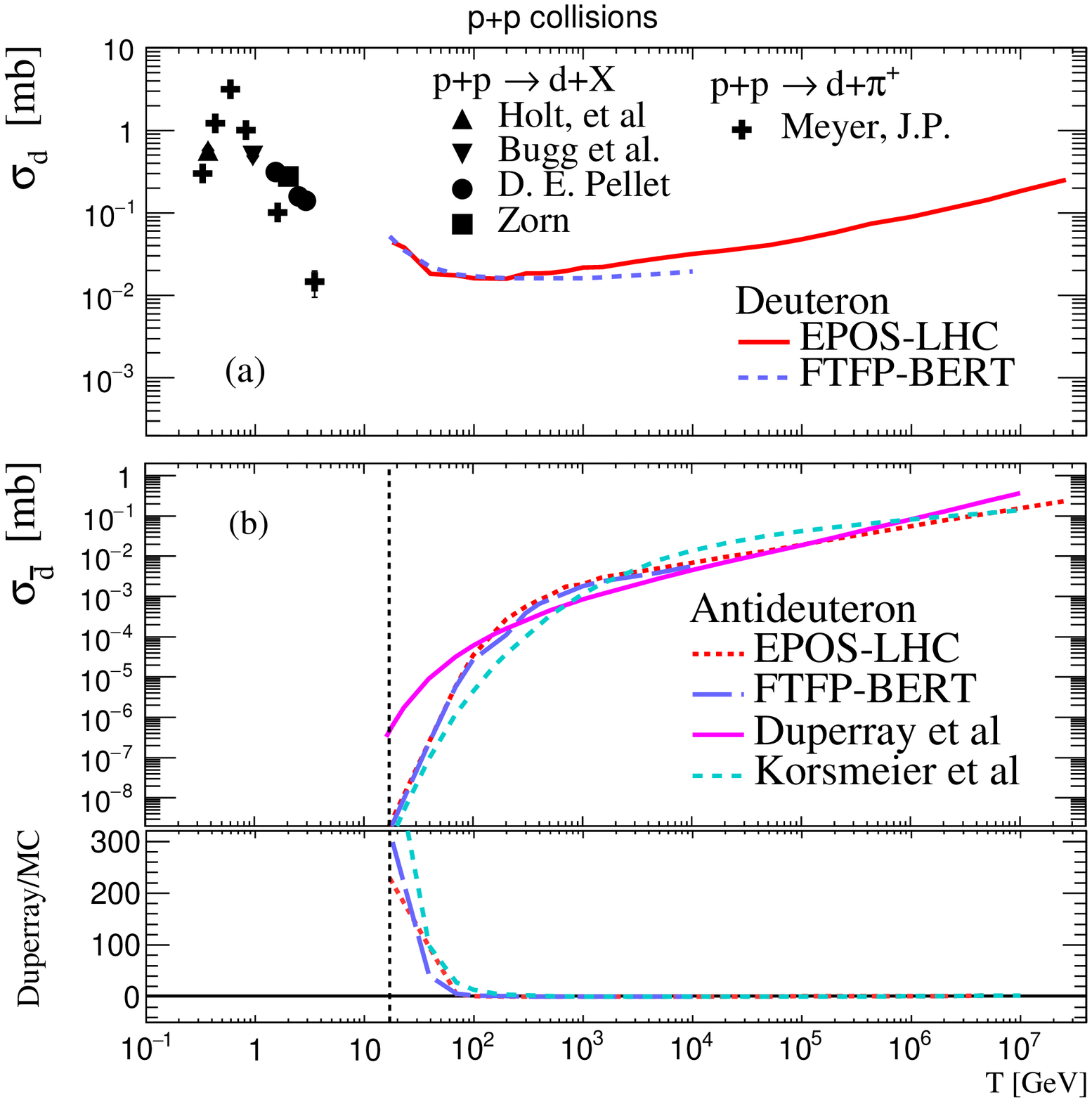}
& \includegraphics[width=8.6cm]{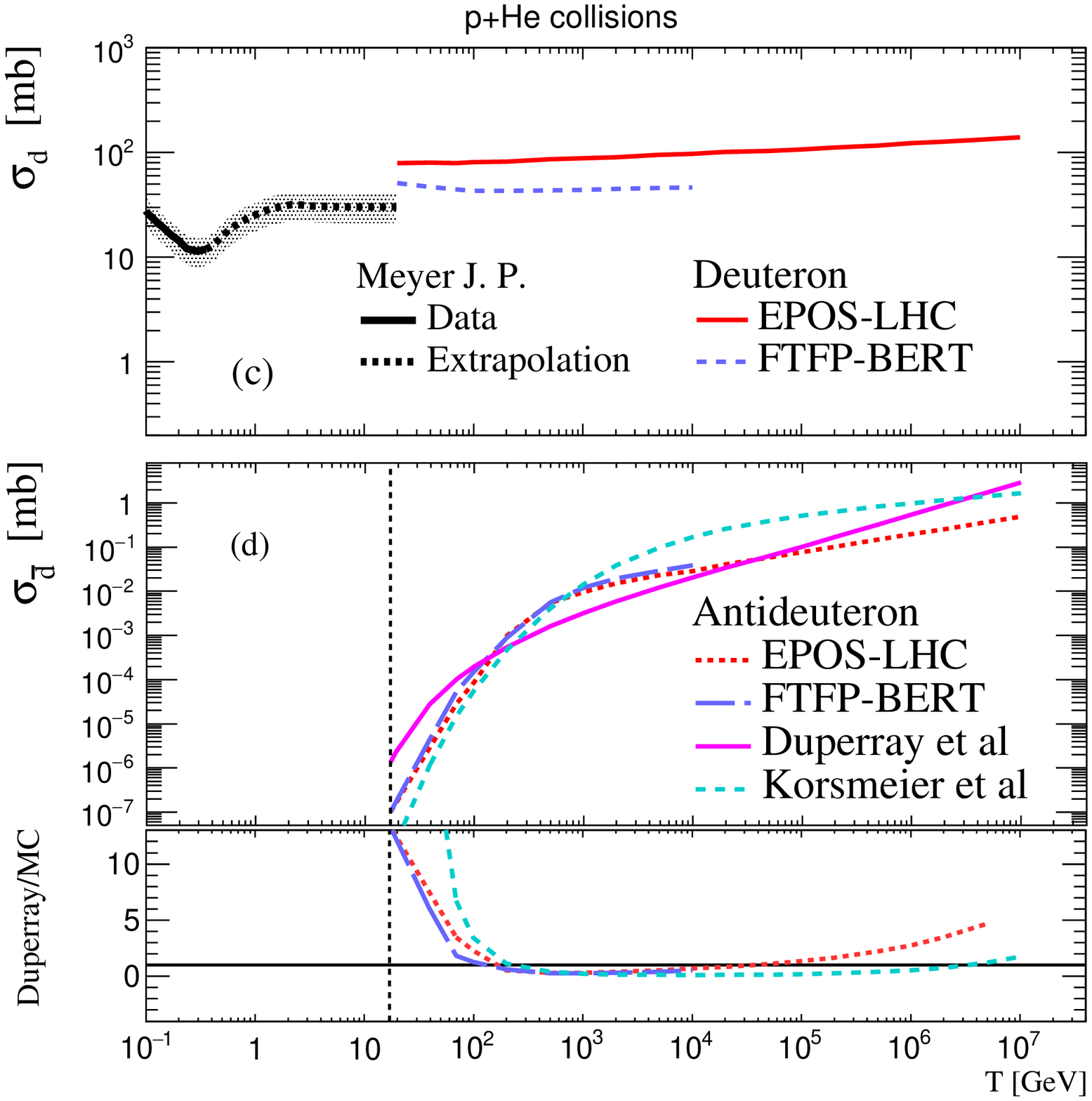}\\
\end{tabular}
\caption{(Color online) Deuteron \textbf{(a)} and antideuteron \textbf{(b)} total production cross section in p+p collisions. Deuteron \textbf{(c)} and antideuteron \textbf{(d)} total production cross section in p+He collisions. The expected antideuteron cross section from Duperray's parametrization has been added. In the lower panels Duperray to MC predictions in antideuteron are compared. Vertical broken lines represent antideuteron production threshold.}
\label{s4:fig6}
\end{center}
\end{figure*}

Note that in the low collision-energy region (T $<$ 100\,GeV) shown in Fig.\,\ref{s4:fig5} (a) the $p_0$ for deuterons decreases reaching a saturation value for T $>$ 100\,GeV. The measurements reported in Table\,\ref{s3:tab1} show that the deuteron production cross section is larger at T $\approx$ 19-24\,GeV than for higher energies. The increase in production seems to be induced by the contribution of opening inelastic channels, not related to coalescence. However, this increase is reproduced in the simulation through the rise in $p_0$ near that particular energy region.

Below 19\,GeV no further comparisons in deuteron production were made, due to limitations of the MC models used. Down at 1$-$3\,GeV, the coalescence model is no longer valid. In this low energy region deuteron production is determined by direct reactions correlated to the initial state as $p+p\rightarrow d+\pi^{+}$, and is independent of similar processes where protons and neutrons are created (as for example $p+p\rightarrow p+n+\pi^{+}$) \cite{Pellett}.

In the case of antideuterons, $p_0$ increases beyond the production threshold (T $\approx$ 17\,GeV) until it saturates at high energies (see Fig.\,\ref{s4:fig5} (b)). Keep in mind that this energy dependence appears in the MC simulations, as well as in the Korsmeier \textit{et al.} parametrization shown in Fig.\,\ref{s4:fig5}, because they reflect best fits to the characteristic trend of the data. However, a gradual growth of $p_0$ beyond the antideuteron production threshold is expected due to phase space \cite{Duperray_2002, Ibarra:2013qt}.

To generate an energy-dependent $p_0$ parameterization that can be used with MC codes, the $p_0$ points shown in Fig.\,\ref{s4:fig5}, have been fitted using Eq.\,\ref{s4:eq1} for deuterons, and Eq.\,\ref{s4:eq2} for antideuterons. The resulting parameters are given in Table\,\ref{s3:tab3}. Since in Fig.\,\ref{s4:fig5} the $p_0$ obtained at certain energy shows no significant differences among p+p, p+Be and p+Al, we used Eq.\,\ref{s4:eq1} and Eq.\,\ref{s4:eq2} to produce a common (target independent) parameterization for deuterons and antideuterons respectively.

Fit function for deuterons:

\begin{equation}\label{s4:eq1}
p_0 = A\left [1+\exp \left (B-\frac{\ln(\text{T/GeV})}{C} \right ) \right ]
\end{equation}

Fit function for antideuterons:

\begin{equation}\label{s4:eq2}
p_0 = \frac{A}{1+\exp(B-\ln(\text{T/GeV})/C)}
\end{equation}

\begin{table}[!htb]
\begin{ruledtabular}
\begin{tabular}{ c  c  c  c }
    \textbf{Model} & \textbf{A} (MeV/$c$) & \textbf{B} & \textbf{C}\\
 \hline 
\multicolumn{4}{c}{} \\
\multicolumn{4}{c}{Deuterons} \\
 \hline
	EPOS-LHC	& 80.6	$\pm$ 2.39 & 4.02 $\pm$ 0.62 & 0.71 $\pm$ 0.11\\
	FTFP-BERT	& 118.1	$\pm$ 2.42 & 5.53 $\pm$ 2.28 & 0.43 $\pm$ 0.14\\	
 \hline      
\multicolumn{4}{c}{} \\
\multicolumn{4}{c}{Antideuterons} \\
 \hline
	EPOS-LHC			& 89.6	$\pm$ 3.0  & 6.6 $\pm$ 0.88 & 0.73 $\pm$ 0.10\\
	FTFP-BERT			& 170.2	$\pm$ 10.5 & 5.8 $\pm$ 0.47 & 0.85 $\pm$ 0.08\\	
	Korsmeier \textit{et al.} \footnotetext[2]{Used with the analytical coalescence model}\footnotemark[2] 	& 153.6 $\pm$ 3.7  & 4.5 $\pm$ 0.36 & 1.47 $\pm$ 0.14\\
\end{tabular}
\end{ruledtabular}
\caption{Values of the parameters for the fitting functions \ref{s4:eq1} and \ref{s4:eq2}.}
\label{s3:tab3}
\end{table} 

\subsection{Total $\text{d}$ and $\bar{\text{d}}$ Production Cross Section}

Based on the coalescence momentum parametrizations of Eq.\,\ref{s4:eq1} and \ref{s4:eq2}, the total deuteron and antideuteron cross sections ($\sigma_{d, \bar{d}}=\sigma_{p+p (p+A)}\times n_{d, \bar{d}}/N_{evt}$) were estimated using the MC simulations to extract the total inelastic cross section ($\sigma_{p+p (p+A)}$), as well as the number of events with at least one $\text{d}$ or $\bar{\text{d}}$ ($n_{d, \bar{d}}$), for a given total number of events ($N_{evt}$). In the Korsmeier \textit{et al.} parametrization case, Eq.\,\ref{s2:eq2} (with antiproton-antineutron independence and symmetry) was integrated using Eq.\,\ref{s4:eq2} and parameters in Table\,\ref{s3:tab3}. The results in p+p and p+He collisions as a function of the collision kinetic energy are plotted in Fig.\,\ref{s4:fig6}, together with available measurements.

The left panels of Fig.\,\ref{s4:fig6} show the results in p+p collisions. The data extracted from Meyer, J. P. \cite{meyer} show the reaction p+p\,$\rightarrow$\,$\text{d}+\pi^{+}$, while the other data \cite{Pellett} and the simulations represent the inclusive reaction p+p\,$\rightarrow$\,$\text{d}+X$. Fig.\,\ref{s4:fig6} (a) shows how deuteron cross section starts to decrease with energy, until it reaches the point-of-inflection of about 100\,GeV which marks the change of slope in the $p_0$ parametrization. From this point, thanks to the constant $p_0$, the cross section starts to grow continuously. The antideuteron cross section on the other hand (Fig.\,\ref{s4:fig6} (b)), emerges from the production threshold and grows rapidly until it changes of slope around T\textasciitilde 1000\,GeV, where the coalescence momentum changes to a constant value. The total antideuteron cross section increases to finally meet the deuteron one at a very high energy.

On the right side of Fig.\,\ref{s4:fig6} the results for p+He collisions are plotted along with data at lower energy from Meyer, J. P. \cite{meyer}. This data only include the reactions: $\text{p}+\text{He}^4$\,$\rightarrow$\,$\text{He}^3+\text{d}$ and $\text{p}+\text{He}^4$\,$\rightarrow$\,$\text{d}+\text{n}+\text{2p}$ (see Fig.\,\ref{s4:fig6} (c)). The simulations have higher values, because they include the coalescence contribution and the fragmentation reactions. However the MC estimation is not far from Meyer extrapolation. The cross section for antideuterons has a similar behavior in p+He as for p+p collisions (see Fig.\,\ref{s4:fig6} (d)), because antinucleons are formed in nucleon-nucleon collisions.

In the lower panels of Figs.\,\ref{s4:fig6} (b) and (d), the ratios of the antideuteron cross section between the Duperray \textit{et al.} parametrization and the results from EPOS-LHC, FTFP-BERT and Korsmeier \textit{et al}. were plotted. As can be observed, the estimations from this work are significantly lower at T$<$100\,GeV than the prediction from Duperray \textit{et al}. This is a direct consequence of the behavior of $p_0$ in this energy region, where instead of having a constant value the coalescence momentum grows gradually.

\section{Conclusions}\label{s-c}

For the purpose of improving the coalescence formation modeling of light nuclei, deuteron and antideuteron production in p+p and p+Be collisions with energies in the laboratory system from 20 to 2.6$\times$ 10$^{7}$\,GeV were reevaluated. As no commonly used hadronic MC generator describes (anti)deuteron production, the goal was to create an afterburner based on experimental data to generate $\text{d}$ and $\bar{\text{d}}$ in p+p and p+A interactions in a reliable way.

After an event-by-event analysis using two of the most relevant MC generators (EPOS-LHC and Geant4's FTFP-BERT), it was found that the coalescence momentum $p_0$ depends on the collision energy (see Fig.\,\ref{s4:fig5}) and is not constant over the entire energy range as previous works suggested. For deuterons, $p_0$ drops with energy until it reaches a constant value, and for antideuterons $p_0$ starts to grow after the production threshold and then reaches a constant value. The behavior of $p_0$ seems to be related with the increase in the available phase space as function of energy \cite{Duperray_2002, Ibarra:2013qt}, however more data in this energy region is necessary to verify this dependence. In addition, it was found there is no substantial difference in the $p_0$ values between p+p and p+Be collisions.

Based on these results parameterizations were developed and used in tandem with EPOS-LHC and FTFP-BERT. Such parameterizations allow us to estimate the differential and total production cross section for deuterons and antideuterons in p+p and p+A collisions (assuming A to be a light nuclei). As an example of the power of this tool, an estimation of the total production cross section of deuterons and antideuterons in p+p and p+He is presented in Fig.\,\ref{s4:fig6}. This new estimation predicts an antideuteron cross section in p+p collisions that can be at least 20 times smaller than the value expected from the parametrization of Duperray \textit{et al.} \cite{Duperray_2003, duperray_flux_2005} in the low kinetic energy (T) region 20-100\,GeV, while at high energies (\textasciitilde 1000\,GeV) the cross section is 2.4 times larger. A similar result is obtained in p+He collisions, where this work estimates a cross section at least 6 times smaller than Duperray \textit{et al.} in the low-T region. Thus, for cosmic-ray applications where a negative power-law describes the energy spectra of the colliding protons, the low-T region is the one that contributes most to the CRs secondary flux, and differences in this area become very important to antideuteron CRs-flux calculations. The detailed quantitative impact of the estimated deuteron and antideuteron production cross sections on the cosmic ray spectra is the subject of an ongoing investigation by our group.\\

\section{Acknowledgments}

The authors would like to thank the scientific computation department of the Institute of Physics, UNAM and to T. Pierog, C. Baus, and R. Ulrich for providing the Cosmic Ray Monte Carlo package. DMGC, AMR and VG would like to thank CONACyT and PAPIIT-DGAPA: IN109617 for the financial support. AD, PVD, and AS would like to thank the National Science Foundation (Award No. 1551980).



\providecommand{\noopsort}[1]{}\providecommand{\singleletter}[1]{#1}%
%


\newpage
\appendix

%
%

\section{Comparison of simulations to accelerator data ($\text{p}$ and $\bar{\text{p}}$)}\label{appendix1}

\begin{figure}
\begin{center}
\includegraphics[width=8.6cm]{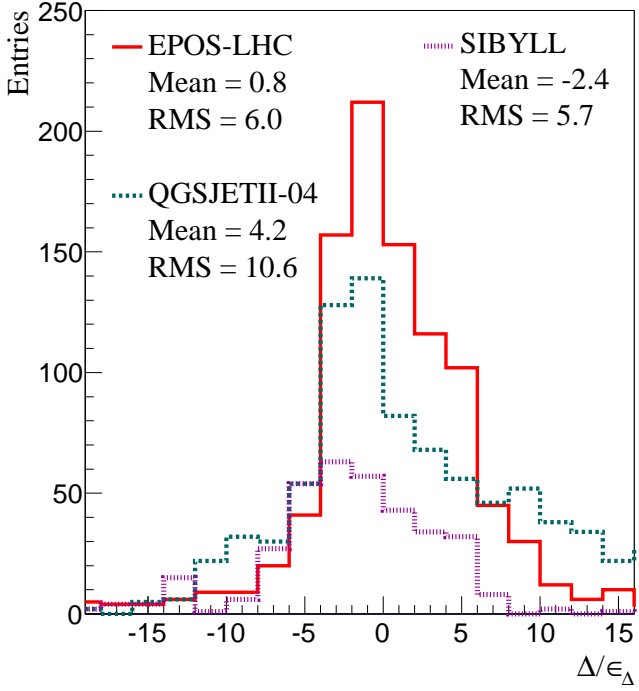}
\caption{(Color online) Distributions of the difference between measurements and the MC generators divided by the error (see Eq.\,\ref{s3:eq1}) for proton production in p+p and p+A collisions.}
\label{ap0:fig1}
\end{center}
\end{figure}

\begin{figure}
\begin{center}
\includegraphics[width=8.6cm]{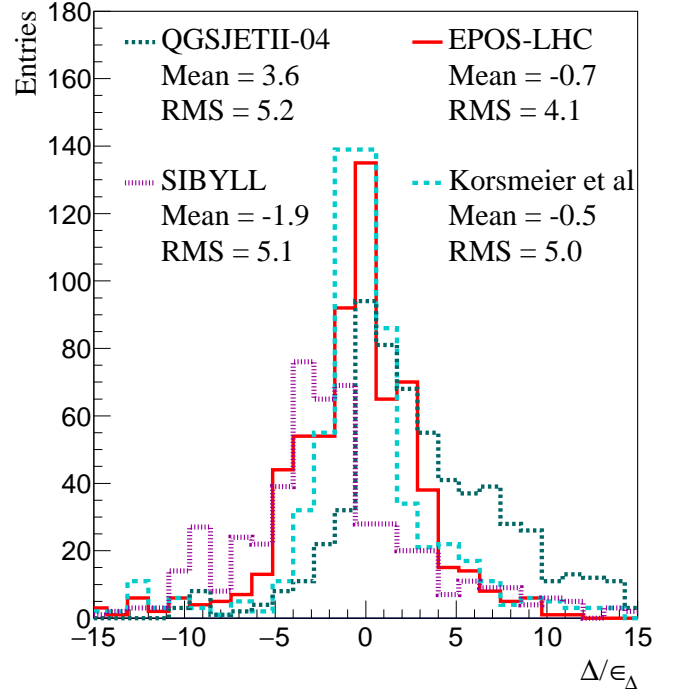}
\caption{(Color online) Distributions of the difference between measurements and the MC generators divided by the error (see Eq.\,\ref{s3:eq1}) for antiproton production in p+p and p+A collisions.}
\label{ap0:fig2}
\end{center}
\end{figure}

Distributions obtained by applying Eq.\,\ref{s3:eq1} to QGSJETII-04 and SIBYLL2.1 are presented and compared with those of EPOS-LHC in Fig.\,\ref{ap0:fig1} for protons and Fig.\,\ref{ap0:fig2} for antiprotons. Fig.\,\ref{ap0:fig2} also includes the parametrization of Korsmeier \textit{et al}.

\begin{figure}
\begin{center}
\includegraphics[width=8.6cm]{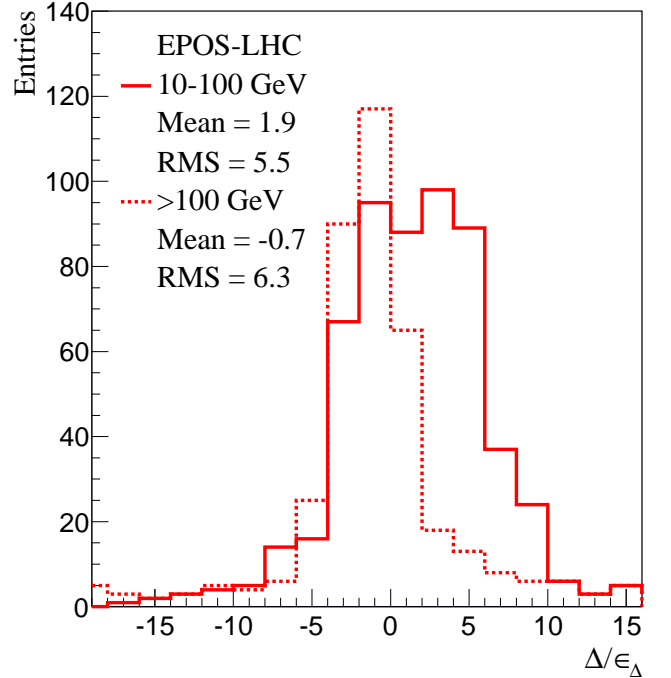}
\caption{(Color online) Distributions in two different energy regions of the difference between measurements and EPOS-LHC divided by the error (see Eq.\,\ref{s3:eq1}) for proton production in p+p and p+A collisions.}
\label{ap0:fig3}
\end{center}
\end{figure}

\begin{figure}
\begin{center}
\includegraphics[width=8.6cm]{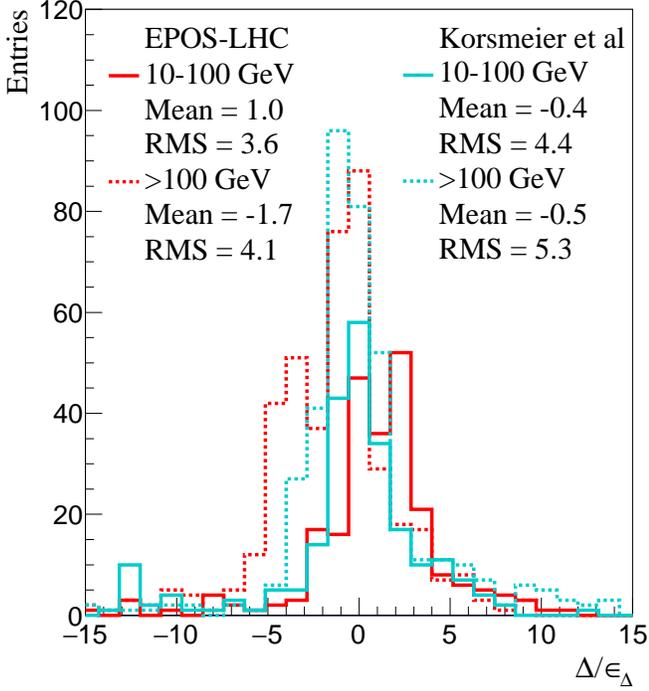}
\caption{(Color online) Distributions in two different energy regions of the difference between measurements and EPOS-LHC divided by the error (see Eq.\,\ref{s3:eq1}) for antiproton production in p+p and p+A collisions.}
\label{ap0:fig4}
\end{center}
\end{figure}

The momenta dependence corresponding to the EPOS-LHC simulation of Fig.\,\ref{s2:fig1} and Fig.\,\ref{s2:fig3} are shown in Fig.\,\ref{ap0:fig3} for protons and Fig.\,\ref{ap0:fig4} for antiprotons. In these plots the distribution was divided in two momentum regions, low (from 10 to 100\,GeV/$c$) and high ($>$ 100\,GeV/$c$). For protons (Fig.\,\ref{ap0:fig3}), the low momentum distribution (solid red line) is shifted to positive values, accounting for the positive value tail in Fig.\,\ref{s2:fig1}. In the high momentum region (dashed red line) the distribution is more symmetric but broader. For antiprotons, the resulting distributions from Korsmeier \textit{et al.} parametrization have also been included in Fig.\,\ref{ap0:fig4}. As can be observed the low momentum distribution of EPOS-LHC is shifted to positive values indicating an overestimation of antiprotons. However, it also shows a lower RMS value compared to the parametrization. The high energy distribution for EPOS-LHC under-predicts antiproton production, revealing that both distributions contribute to the positive and negative value tails in Fig.\,\ref{s2:fig3}.

%
%

\section{Comparison of simulations to accelerator data ($\text{p}$, $\bar{\text{p}}$, $\text{d}$ and $\bar{\text{d}}$)}\label{appendix}

This appendix is a collection of all comparisons made between accelerator data and MC models. The three MC models studied are plotted in each figure with the same marker and color convention: EPOS-LHC (red circle \tikzcircle{3pt}); FTFP-BERT (blue square \tikzsquare{blue}); and QGSP-BERT (green triangle \tikztriangle{black!30!green}). Data are presented as black dots or black squares. The comparisons are shown for either the differential cross sections or invariant differential cross sections as a function of laboratory or transverse momentum per nucleon. When possible, (anti)protons and (anti)deuterons are shown in the same figure.

\subsection{p+p and p+Be at $\mathbf{p_{lab} = 19.2}$\,GeV/$\mathbf{c}$}

\begin{figure}
\begin{center}
\includegraphics[width=8.6cm]{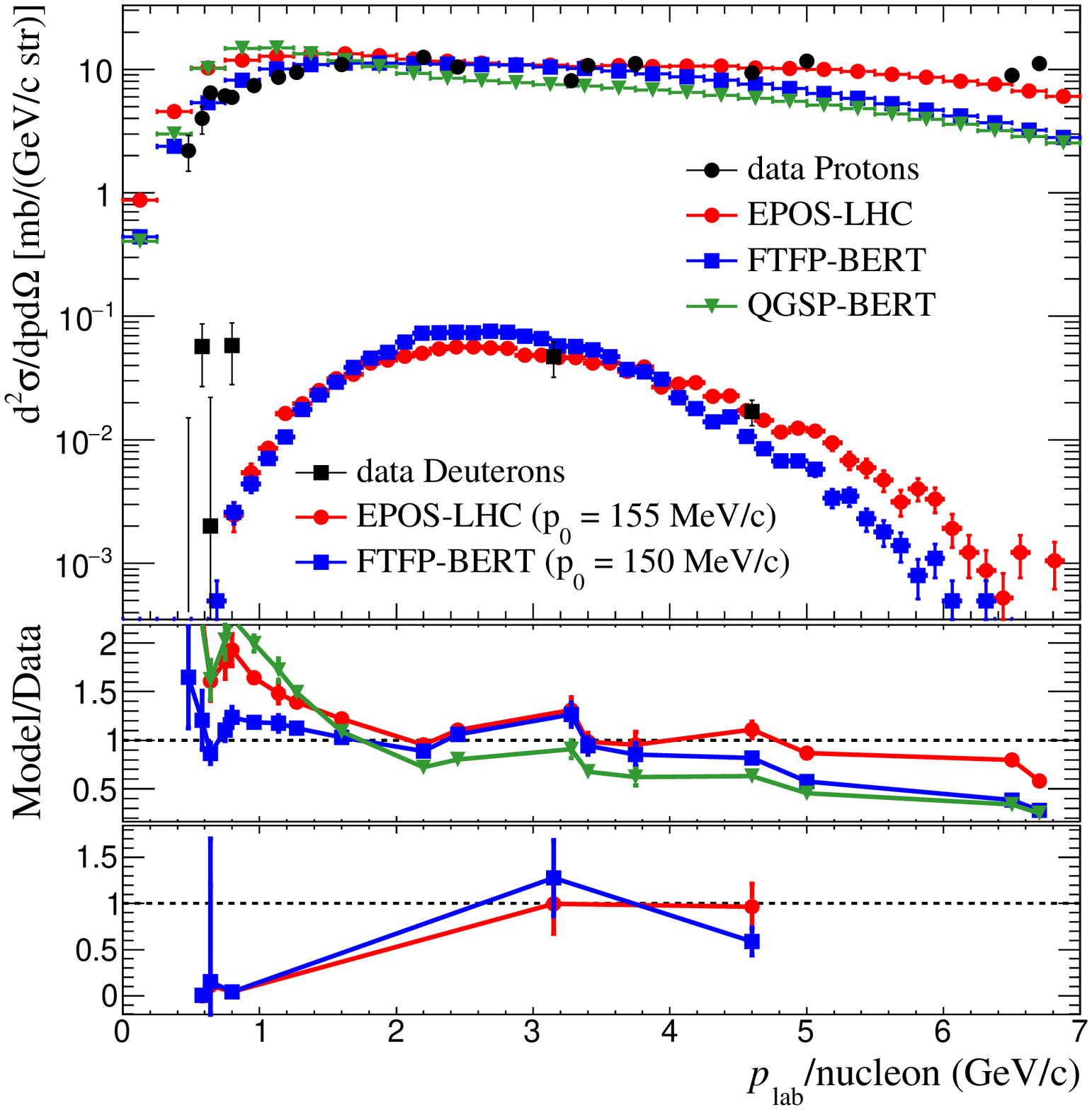}
\caption{Double differential cross sections from MC models compared to data of protons and deuterons produced in p+p collisions at 19\,GeV/$c$ \cite{diddens_particle_2008}.}
\label{ap1:fig1}
\end{center}
\end{figure} 

\begin{figure}
\begin{center}
\includegraphics[width=8.6cm]{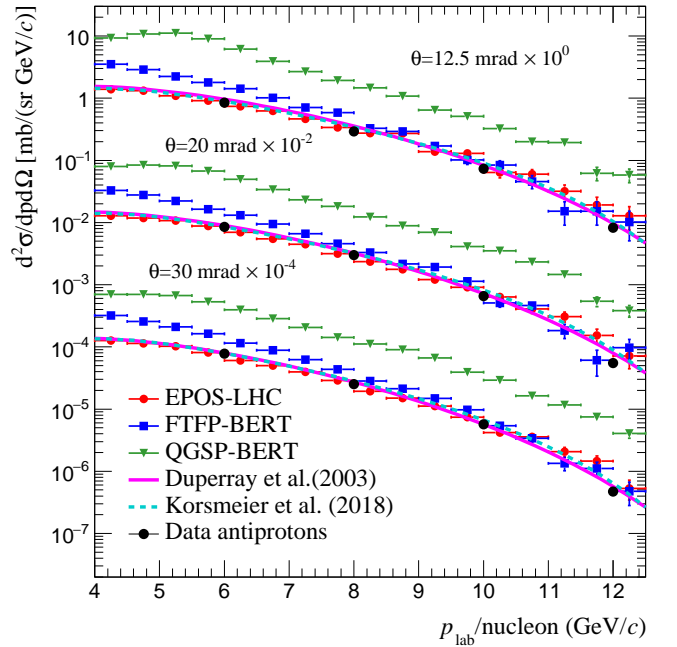}
\caption{Double differential cross sections from MC models and Duperray's parametrization (pink line) compared to data of antiprotons produced in p+Be collisions at 19.2\,GeV/$c$ \cite{Allaby1970}.}
\label{ap1:fig2}
\end{center}
\end{figure} 

Results from \cite{Allaby1970} show p and $\bar{\text{p}}$ production in p+p, p+Be and p+Al collisions. The nucleons produced cover a laboratory momentum range from 2 to 19\,GeV and an angular region from 12.5 to 70\,mrad. Another experiment \cite{diddens_particle_2008} at nearly the same energy (19\,GeV/$c$) reported p, $\bar{\text{p}}$ and d production in p+p collisions for $\theta$ = 116\,mrad.  

In Fig.\,\ref{ap1:fig1}, proton and deuteron production in p+p are shown in comparison to data of \cite{diddens_particle_2008}. Values of $p_0$ = 155\,MeV/$c$ and $p_0$ = 150\,MeV/$c$ were determined from the fit to deuteron data with EPOS-LHC and FTFP-BERT, respectively. In Fig.\,\ref{ap1:fig2}, antiproton production in p+Be collisions is shown for three different angles, alongside with the parameterization of Duperray \cite{Duperray_2003} (magenta continuous line).

\begin{figure}[!htb]
\begin{center}
\includegraphics[width=8.6cm]{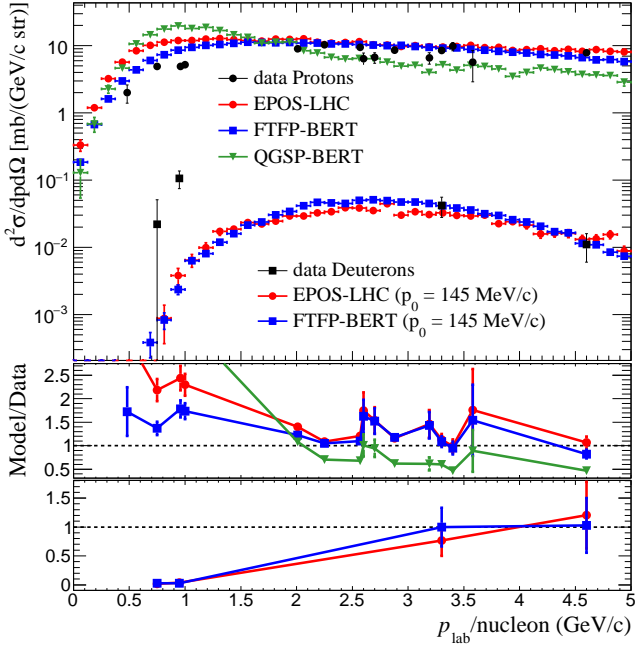}
\caption{Double differential cross sections from MC models compared to data of protons and deuterons produced in p+p collisions at 24\,GeV/$c$ \cite{diddens_particle_2008}.}
\label{ap1:fig3}
\end{center}
\end{figure} 

\begin{figure}[!htb]
\begin{center}
\includegraphics[width=8.6cm]{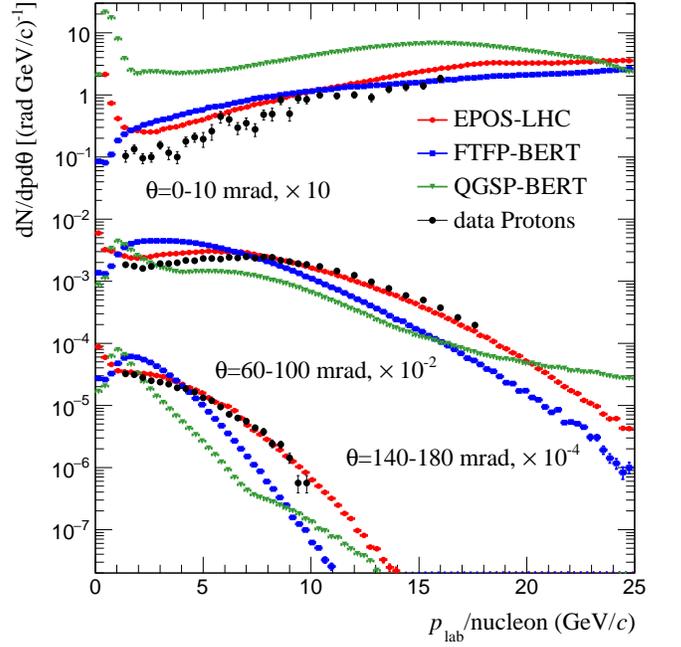}
\caption{Double differential momentum distribution from MC models compared to data of protons produced in p+C collisions at 31\,GeV/$c$ \cite{Abgrall2016}.}
\label{ap1:fig4}
\end{center}
\end{figure} 

\subsection{p+p at $\mathbf{p_{lab} = 24}$\,GeV/$\mathbf{c}$}
The same group that measured p, $\bar{\text{p}}$ and d production in p+p collisions at 19\,GeV also reported results at 24\,GeV \cite{diddens_particle_2008}. The results are compared with the MC models in Fig.~\ref{ap1:fig3}. Best fit values of the coalescence momentum for deuterons are $p_0 = 145$\,MeV/$c$ and $p_0 = 145$\,MeV/$c$ for EPOS-LHC and FTFP-BERT.

\subsection{p+C at $\mathbf{p_{lab} = 31}$\,GeV/$\mathbf{c}$}

The NA61/SHINE collaboration reported the production of mesons and baryons in p+C collisions at an incoming momentum of 31\,GeV/$c$ in 2016 \cite{Abgrall2016}. In Fig.~\ref{ap1:fig4} data at three different angles is plotted in comparison with MC models. 

\begin{figure}[!htb]
\begin{center}
\includegraphics[width=8.6cm]{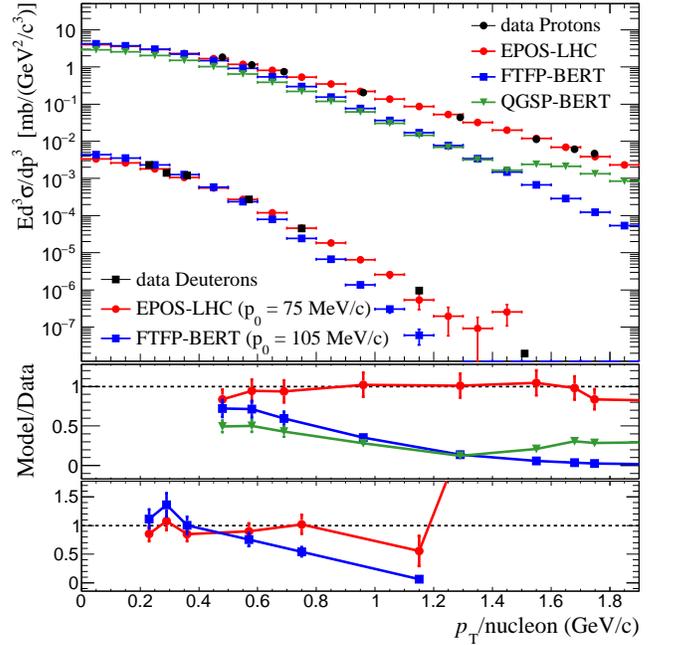}
\caption{Invariant differential cross section for protons and deuterons produced in p+p collisions at 70\,GeV/$c$. Data taken from \cite{abramov_production_1980, Abramov_Baldin1985, Abramov_Baldin1987}.}
\label{ap1:fig5}
\end{center}
\end{figure} 

\subsection{p+p, p+Be and p+Al at $\mathbf{p_{lab} = 70}$\,GeV/$\mathbf{c}$}

\begin{figure}[!htb]
\begin{center}
\includegraphics[width=8.6cm]{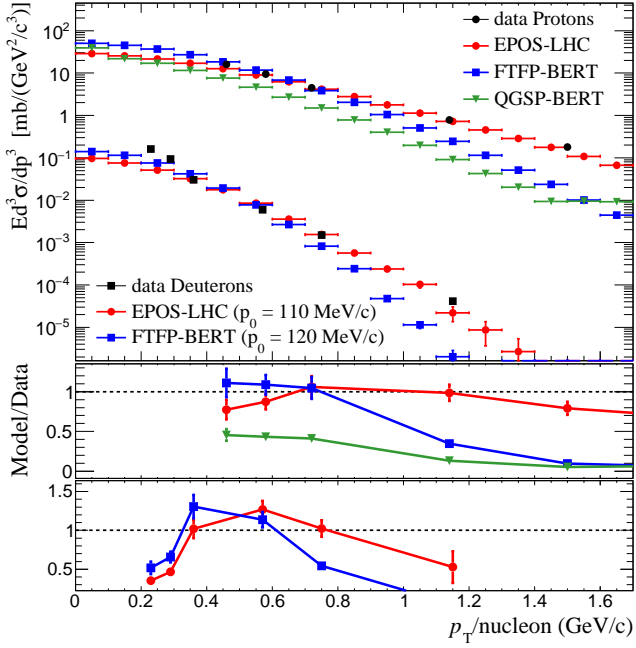}
\caption{Invariant differential cross section for protons and deuteron produced in p+Be collisions at 70\,GeV/$c$. Data taken from \cite{Abramov_Baldin1987, Abramov_Baldin1987}.}
\label{ap1:fig7}
\end{center}
\end{figure}

\begin{figure}[!htb]
\begin{center}
\includegraphics[width=8.6cm]{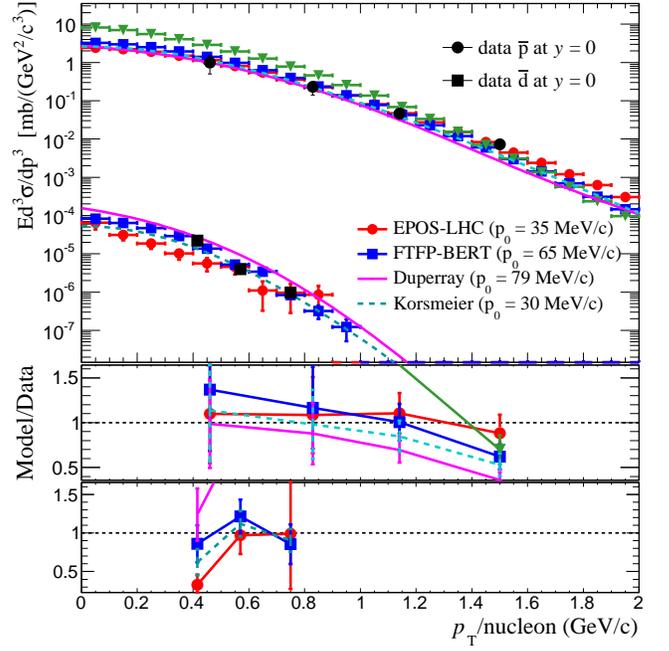}
\caption{Invariant differential cross section for antiprotons and antideuterons produced in p+Be collisions at 70\,GeV/$c$. Data taken from \cite{Abramov_Baldin1987, Abramov_Baldin1987}.}
\label{ap1:fig8}
\end{center}
\end{figure}

A series of experiments performed in the Russian Institute for High Energy Physics at Serpukhov measured the production of p, $\bar{\text{p}}$, d and $\bar{\text{d}}$ in p+p, p+Be and p+Al collisions at 70\,GeV/$c$ \cite{abramov_production_1980, Abramov_Baldin1985, Abramov_Baldin1987, Abramov:1983es}. Protons and antiprotons were detected in a transverse momentum region from 0.48 to 4.22\,GeV/$c$ and deuterons and antideuterons were evaluated until $p_{T}\approx$ 3.8\,GeV/$c$. Both hadrons and nuclei were measured at an angle of $\theta = 160$\,mrad or 90$^{\circ}$ in the center-of-mass frame. Figs.\,\ref{ap1:fig5}, \ref{s4:fig4}, \ref{ap1:fig7} and \ref{ap1:fig8} present this set of data in comparison with MC generators. The best fit values for $p_0$ are shown in the figures. Despite the fact that some authors like Duperray \textit{et al.} \cite{Duperray_2003, duperray_flux_2005} excluded these data from their analysis, the authors of this study did not find a reason to reject them. Besides, this is the lowest energy at which the spectrum of the invariant antideuteron cross section was measured so far.

\subsection{p+p, p+C at $\mathbf{p_{lab} = 158}$\,GeV/$\mathbf{c}$}

\begin{figure}[!htb]
\begin{center}
\includegraphics[width=8.6cm]{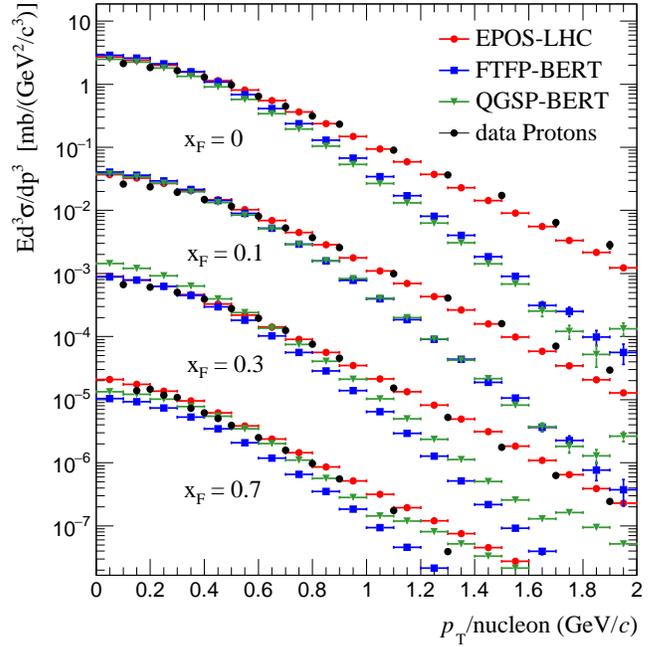}
\caption{Invariant differential cross section for protons produced in p+p collisions at 158\,GeV/$c$. Data taken from \cite{Anticic:2009wd}.}
\label{ap1:fig9}
\end{center}
\end{figure}

\begin{figure}[!htb]
\begin{center}
\includegraphics[width=8.6cm]{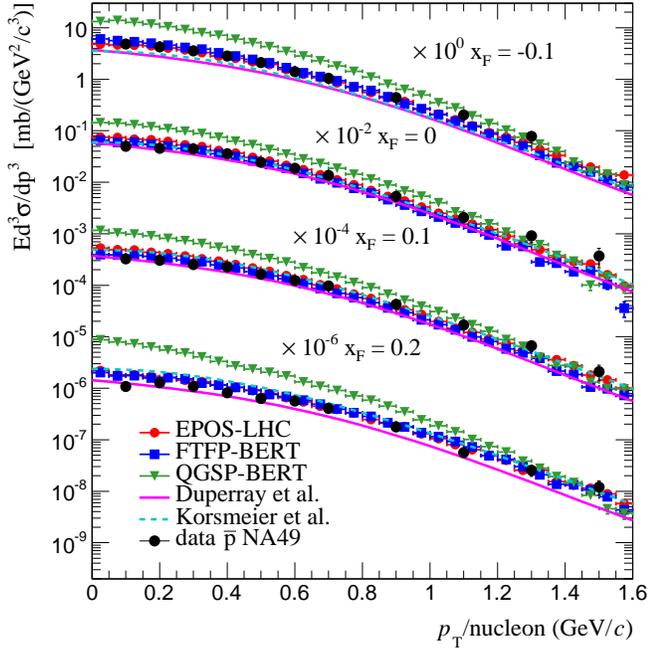}
\caption{Invariant differential cross section for antiprotons produced in p+C collisions at 158\,GeV/$c$. Data taken from \cite{Baatar:2012fua}.}
\label{ap1:fig10}
\end{center}
\end{figure}

NA49 experiment published results on the production of protons, deuterons and antiprotons in p+p and p+C collisions at 158\,GeV/$c$ in 2009 and 2012 \cite{Anticic:2009wd, Baatar:2012fua}. These modern data sets are important since they are achieved with up-to-date techniques in hardware and data analysis and have low systematic errors. Figs.\,\ref{ap1:fig9} and \ref{ap1:fig10} show the invariant differential cross sections as function of $p_T$ for different values of Feynman $x_F$ calculated with MC and compared with data. Only protons from p+p collisions (Fig.\,\ref{ap1:fig9}) and antiprotons from p+C collisions (Fig.\,\ref{ap1:fig10}) are displayed, however, the analysis also includes antiprotons from p+p and protons from p+C. 

\subsection{p+Be, p+Al at $\mathbf{p_{lab} = 200}$\,GeV/$\mathbf{c}$}

\begin{figure}[!htb]
\begin{center}
\includegraphics[width=8.6cm]{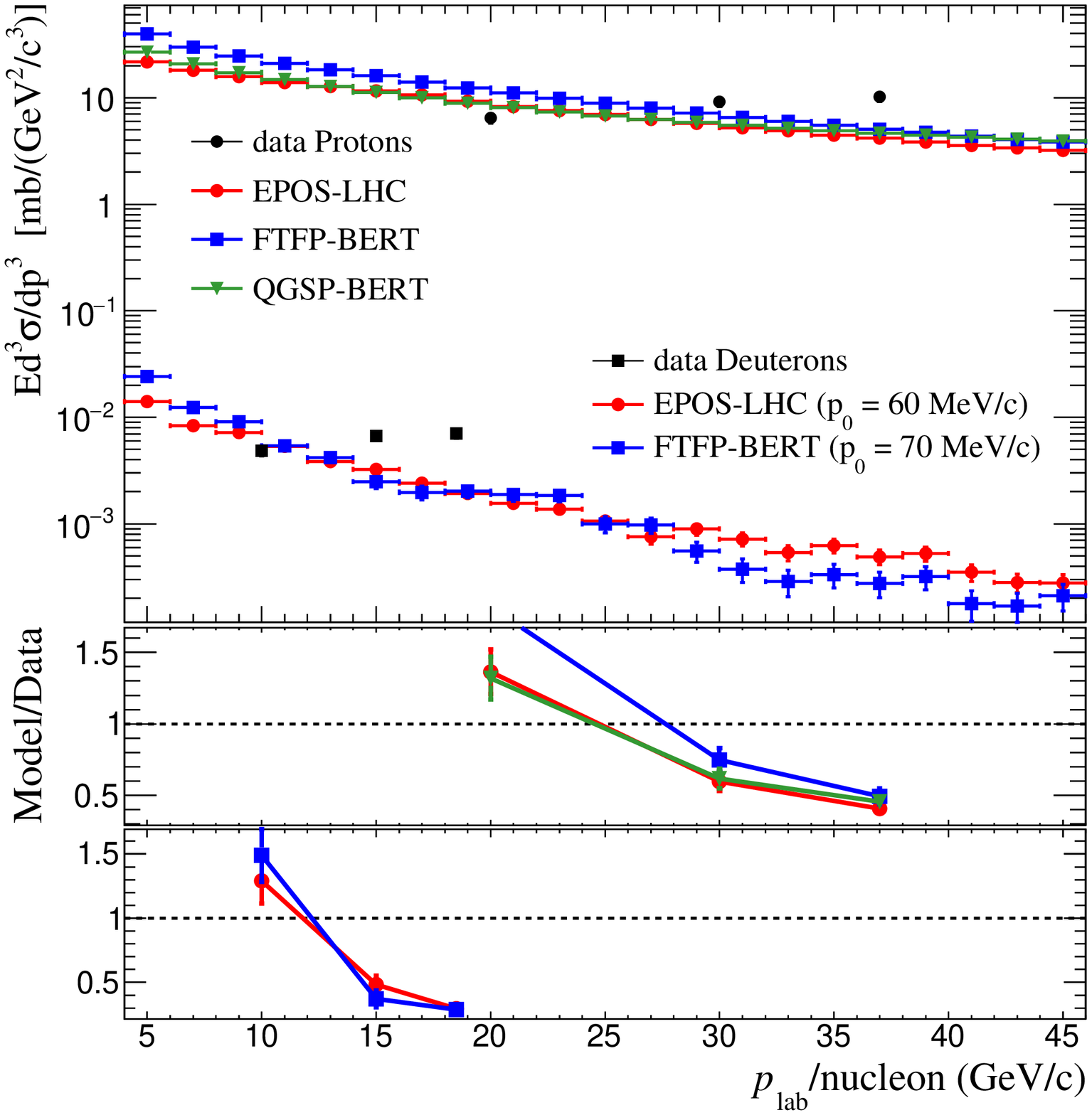}
\caption{Invariant differential cross section for protons and deuteron produced in p+Be collisions at 200\,GeV/$c$. Data taken from \cite{BOZZOLI1978317, Bussiere:1980yq}.}
\label{ap1:fig11}
\end{center}
\end{figure}

\begin{figure}[!htb]
\begin{center}
\includegraphics[width=8.6cm]{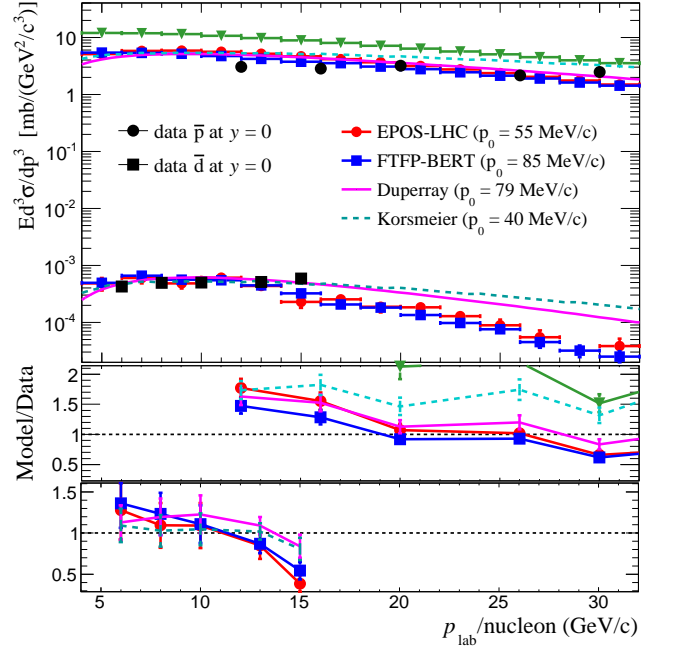}
\caption{Invariant differential cross section for antiprotons and antideuterons produced in p+Be collisions at 200\,GeV/$c$. Data taken from \cite{BOZZOLI1978317, Bussiere:1980yq}.}
\label{ap1:fig12}
\end{center}
\end{figure}

Protons, antiprotons, deuterons, and antideuterons produced in p+Be and p+Al collisions using the CERN-SPS accelerator were measured by \cite{BOZZOLI1978317, Bussiere:1980yq}. Proton and antiproton production was also measured at the Fermi National Accelerator Laboratory between 23\,GeV/$c$ and 200\,GeV/$c$ in p+Be collisions at 3.6\,mrad \cite{Baker:1974fv}. Data from CERN were reported as ratios of differential cross section with respect to pions. Following the procedure used by \cite{Duperray_2003}, the differential cross sections were calculated from the measured ratios. Results in p+Be for protons and deuterons are presented in Fig.\,\ref{ap1:fig11} while results for antiprotons and antideuterons are shown in Fig.\,\ref{ap1:fig12}. 

\subsection{p+p, p+Be at $\mathbf{p_{lab} = 300}$~and~$\mathbf{400}$\,GeV/$\mathbf{c}$}

\begin{figure}[!htb]
\begin{center}
\includegraphics[width=8.6cm]{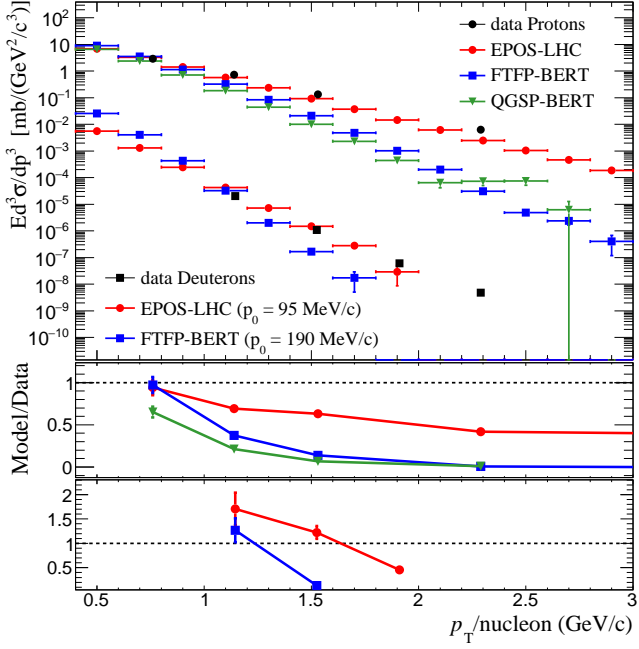}
\caption{Invariant differential cross section for protons and deuterons produced in p+Be collisions at 300\,GeV/$c$. Data taken from \cite{Antreasyan, Cronin}.}
\label{ap1:fig13}
\end{center}
\end{figure}

\begin{figure}[!htb]
\begin{center}
\includegraphics[width=8.6cm]{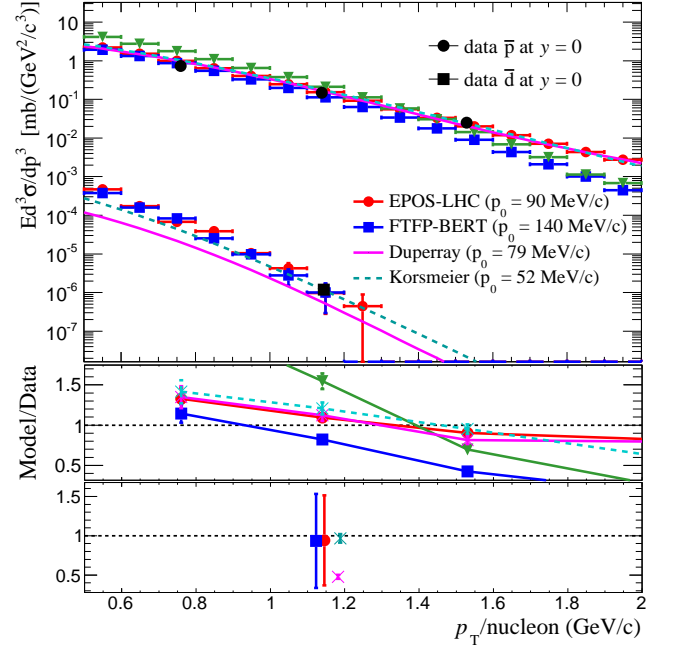}
\caption{Invariant differential cross section for antiprotons and antideuterons produced in p+Be collisions at 300\,GeV/$c$. Data taken from \cite{Antreasyan, Cronin}.}
\label{ap1:fig14}
\end{center}
\end{figure}

A large group of measurements were conducted at the Fermilab synchrotron with incident momenta of 200, 300 and 400\,GeV/$c$ using various targets, such as p, $\text{D}_2$, Be, Ti and W. Protons and antiprotons were measured for every type of collision, but deuterons and antideuterons were only extracted at 300\,GeV/$c$ and measured at large transverse momentum $p_T/\text{nucleon} > 1$\,GeV/$c$. All the particles emitted from collisions were computed at 77\,mrad which corresponds to an angle of $\approx$ 90$^{\circ}$ in the center-of-mass system \cite{Antreasyan, Cronin}. The specific case of p+Be at 300\,GeV/$c$ compared to MC models is shown in Figs.\,\ref{ap1:fig13} and \ref{ap1:fig14}.  

\subsection{p+p at $\mathbf{\sqrt{s} = 45}$~and~$\mathbf{53}$\,GeV}

\begin{figure}[!htb]
\begin{center}
\includegraphics[width=8.6cm]{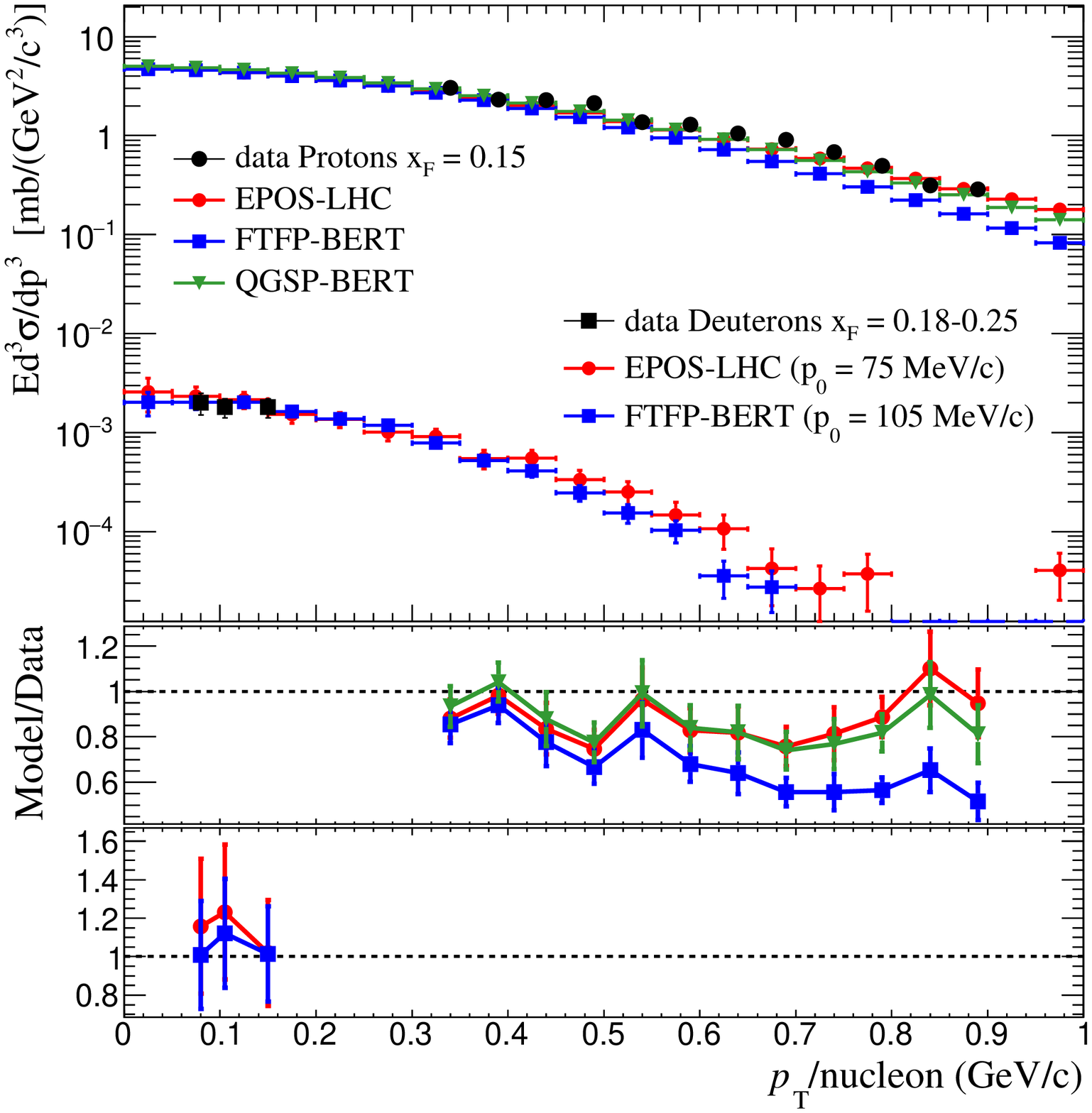}
\caption{Invariant differential cross section for protons and deuteron produced in p+p collisions at $\sqrt{s}$ = 53\,GeV. Data taken from \cite{Alper1975, albrow}.}
\label{ap1:fig15}
\end{center}
\end{figure}

\begin{figure}[!htb]
\begin{center}
\includegraphics[width=8.6cm]{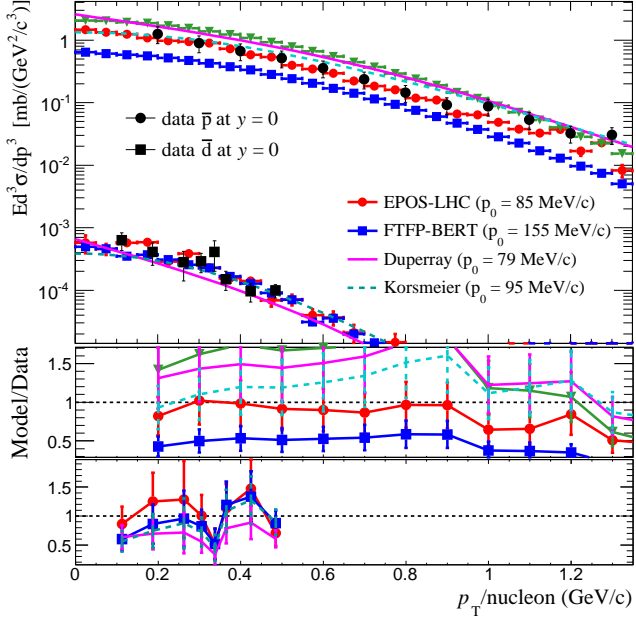}
\caption{Invariant differential cross section for antiprotons and antideuterons produced in p+p collisions at $\sqrt{s}$ = 53\,GeV. Data taken from \cite{Alper1975, Alper2, gibson_production_2008}.}
\label{ap1:fig16}
\end{center}
\end{figure}

The production of pions, kaons, nucleons and antinucleons was measured at the CERN Intersecting Storage Ring in p+p collisions at a variety of energies in the center-of-mass frame with $\sqrt{s}=$ 23, 31, 45, 53, 63\,GeV \cite{Alper1975}. Deuterons and antideuterons were only reported for 45 and 53\,GeV \cite{Alper2, gibson_production_2008, albrow}. Following the analysis of proton and antiproton production by the NA49 collaboration, a feed down excess of 25\% was estimated from simulations and it was applied to the whole sample. This correction significantly reduces the proton production, but leaves antiprotons essentially unchanged because of systematic errors in the nuclear absorption correction of about 30\%. Results are shown in Figs.\,\ref{ap1:fig15} and \ref{ap1:fig16}.

\subsection{p+He at $\mathbf{\sqrt{s_{NN}} = 110}$\,GeV}
\begin{figure}[!htb]
\begin{center}
\includegraphics[width=8.6cm]{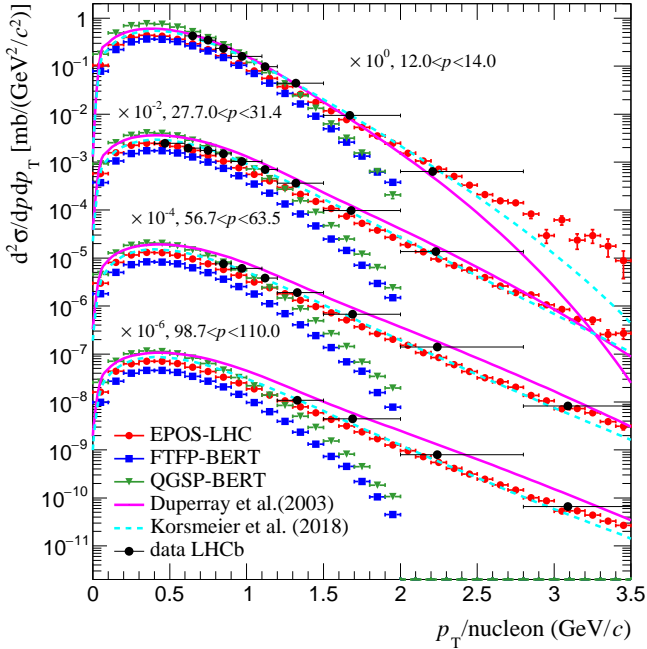}
\caption{Differential cross section for antiprotons produced in p+He collisions at $\sqrt{s_{NN}}$ = 110\,GeV. Data taken from \cite{lhcb}.}
\label{ap1:fig17}
\end{center}
\end{figure}

Antiprotons produced in p+He collisions with a 6.5\,TeV proton beam were measured recently by the LHCb experiment at CERN. The antiproton momentum range covered was from 12 to 110\,GeV/c. The antiprotons collected were produced only by direct collisions or from resonances decaying via strong interaction. In Fig.\,\ref{ap1:fig17} the data is compared with the MC models EPOS-LHC, FTFP-BERT, and QGSP-BERT. The parametrizations from Duperray and Korsmeier are also included.

\subsection{p+p at $\mathbf{\sqrt{s} = 900}$~and~$\mathbf{7000}$\,GeV}
\begin{figure}[!htb]
\begin{center}
\includegraphics[width=8.6cm]{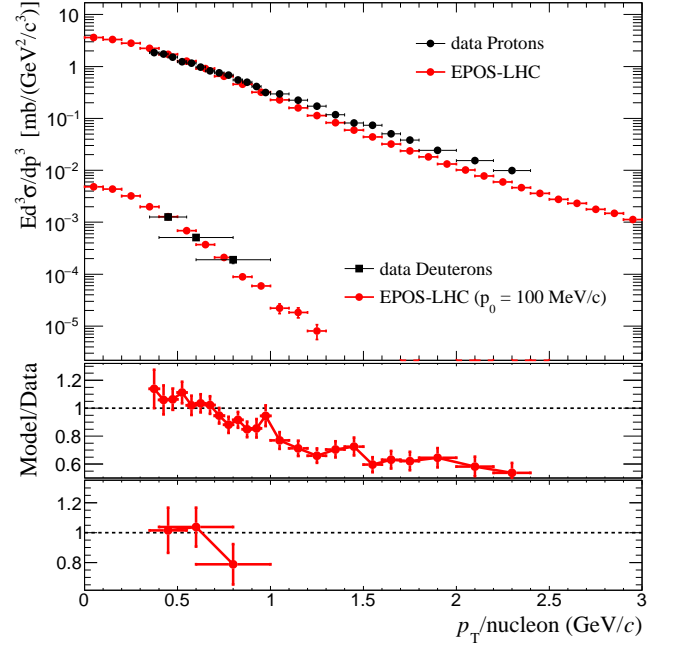}
\caption{Invariant differential cross section for protons and deuteron produced in p+p collisions at $\sqrt{s}$ = 900\,GeV. Data taken from \cite{aamodt_production_2011, eulogio, eulogio_paper}.}
\label{ap1:fig18}
\end{center}
\end{figure}

\begin{figure}[!htb]
\begin{center}
\includegraphics[width=8.6cm]{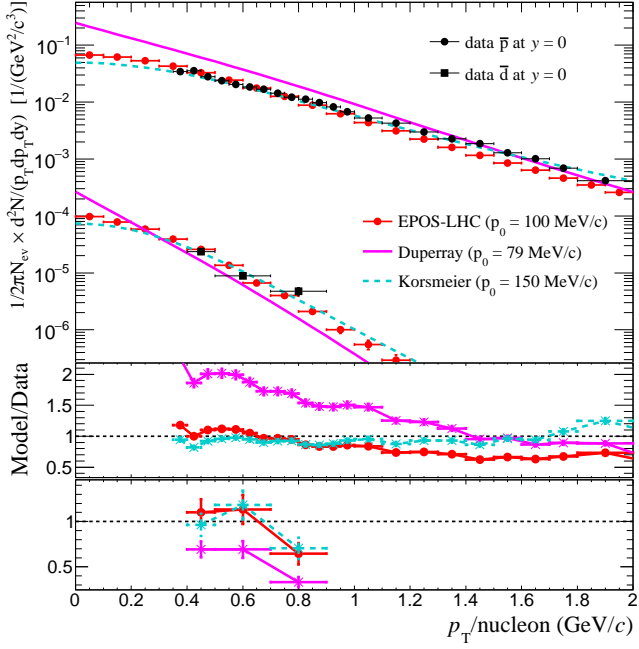}
\caption{Invariant differential momentum distribution for antiprotons and antideuterons produced in p+p collisions at $\sqrt{s}$ = 900\,GeV. Data taken from \cite{aamodt_production_2011, eulogio, eulogio_paper}.}
\label{ap1:fig19}
\end{center}
\end{figure}

At the LHC, protons and antiprotons as well as deuterons and antideuterons are produced in p+p and Pb+Pb collisions at very high energies. ALICE reported results at 0.9, 2.76 and 7\,TeV in the central rapidity region -0.5 $<y<$ 0.5 for a wide range of transverse momentum ($p_T<$ 5\,GeV/$c$) \cite{aamodt_production_2011, eulogio, eulogio_paper, alicedeuteron}. The data are compared with EPOS-LHC and the Duperray parameterization in Figs.\,\ref{ap1:fig18} and \ref{ap1:fig19}. FTFP and QGSP were not included, since Geant4 models have an energy limit of $\sqrt{s} \approx$ 430\,GeV.

%
%

\section{(Anti)proton mismatch factorization for EPOS-LHC and FTFP-BERT}\label{appendix2}

Assuming (anti)proton-(anti)neutron independence and symmetry, Eq.\,\ref{s2:eq2} can be rewritten as:

\begin{equation}\label{s4:eq3}
\gamma_{\bar{d}} \frac{d N_{\bar{d}}}{d\vec{k}_{\bar{d}}^3}^{sim}(\vec{k}_{\bar{d}}) = \frac{4\pi p_0^3}{3} \left ( \gamma_{\bar{p}} \frac{d N_{\bar{p}}}{d\vec{k}_{\bar{p}}^3}^{sim}(\vec{k}_{\bar{p}}) \right )^2
\end{equation}

The proton or antiproton mismatch can be represented by the energy-dependent ratio.

\begin{equation}
r\text{(T)}=\left ( \frac{\gamma_{\bar{d}}\frac{d N_{\bar{p}}}{d\vec{k}_{\bar{p}}^3}^{sim}}{\gamma_{\bar{p}}\frac{d N_{\bar{p}}}{d\vec{k}_{\bar{p}}^3}^{data}} \right ). 
\end{equation}

Inserting the $r(\text{T})$ factor in Eq.\,\ref{s4:eq3}, the final result is:

\begin{equation}\label{s4:eq4}
\gamma_{\bar{d}} \frac{d N_{\bar{d}}}{d\vec{k}_{\bar{d}}^3}^{sim}(\vec{k}_{\bar{d}}) = \frac{4\pi}{3} (p^{\prime}_{0})^{3} \left ( \gamma_{\bar{p}} \frac{d N_{\bar{p}}}{d\vec{k}_{\bar{p}}^3}^{data}(\vec{k}_{\bar{p}}) \right )^2
\end{equation}

\begin{figure*}[!h]
\begin{center}
\begin{tabular}{ll}
\includegraphics[width=8.6cm]{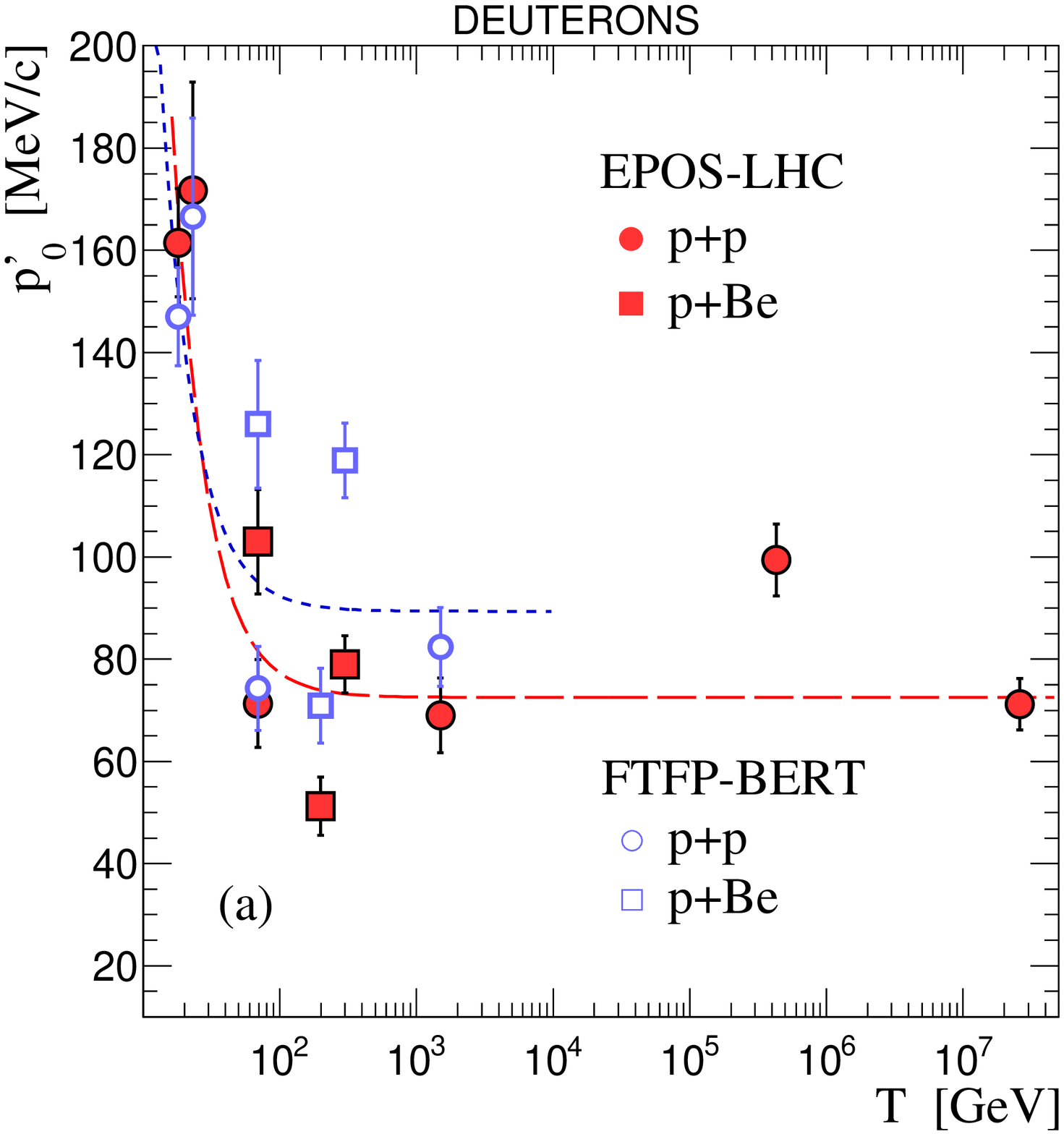}
& \includegraphics[width=8.6cm]{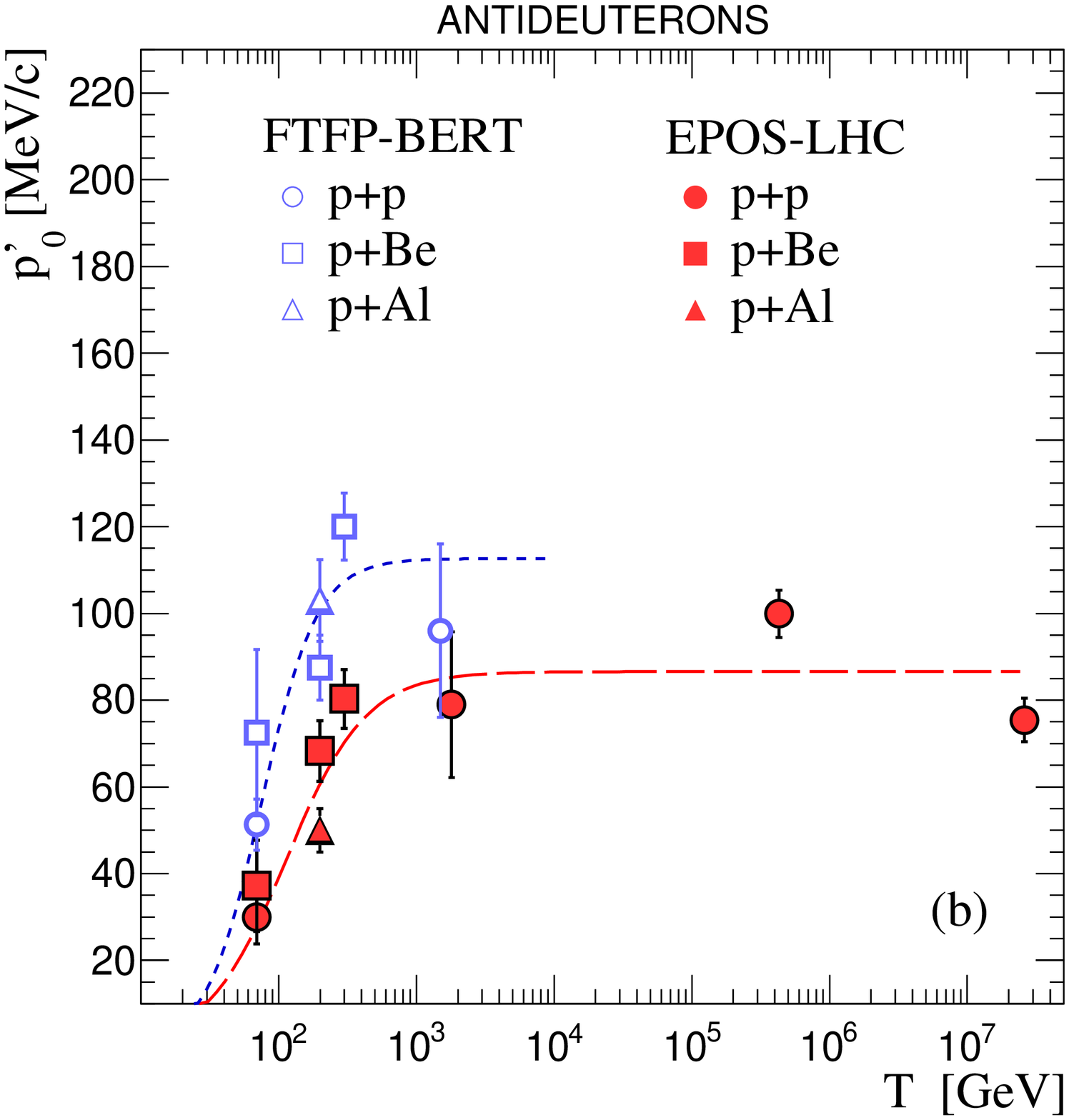}\\
\end{tabular}
\caption{(Color online) Extracted coalescence momentum $p^{\prime}_{0}$ (symbols) for deuterons \textbf{(a)} and  antideuterons \textbf{(b)} as function of the collision kinetic energy (T). Fit functions [Eqs.\,(\ref{s4:eq1}) and (\ref{s4:eq2})] for EPOS-LHC (long-dashed red line) and FTFP-BERT (dashed blue line) are shown.}
\label{ap2:fig1}
\end{center}
\end{figure*}

Where $p^{\prime}_{0}=p_0\cdot r(\text{T})^{2/3}$, is the redefined coalescence momentum that is now more specific to the coalescence process rather than scaling the mismatch of the (anti)protons. The values of $p^{\prime}_{0}$ for EPOS-LHC and FTFP-BERT are shown in Fig.\,\ref{ap2:fig1} as function of the collision kinetic energy (T). As observed, after factorizing the mismatch the $p^{\prime}_{0}$ values of FTFP-BERT are close to the values of EPOS-LHC, showing a similar energy dependence. This, justified the use of Eqs. \ref{s4:eq1} and \ref{s4:eq2} to fit the extracted $p_0$ for both models. Differences in $p^{\prime}_{0}$ for EPOS-LHC and FTFP-BERT after the mismatch factorization, are related to the intrinsic effects of the models as for example (anti)nucleon production asymmetries. 

\onecolumngrid

\twocolumngrid
 
\end{document}